\documentclass[preprint2]{aastex}

\newcommand{\HII}{H\,{\sc ii}}
\newcommand{\NII}{[N\,{\sc ii}]}
\newcommand{\OII}{[O\,{\sc ii}]}
\newcommand{\OIII}{[O\,{\sc iii}]}
\newcommand{\SII}{[O\,{\sc ii}]}

\newcommand{\Ha}{H$\alpha$}
\newcommand{\Hb}{H$\beta$}

\newcommand{\abox}{12+log(O/H)}

\newcommand{\Ms}{$M_{\star}$}

\newcommand{\Te}{$T_{\rm e}$}

\DeclareRobustCommand{\ion}[2]{%
\relax\ifmmode
\ifx\testbx\f@series
{\mathbf{#1\,\mathsc{#2}}}\else
{\mathrm{#1\,\mathsc{#2}}}\fi
\else\textup{#1\,{\mdseries\textsc{#2}}}%
\fi}

\usepackage{graphicx} 
\usepackage[dvips]{epsfig}
\usepackage{txfonts}
\usepackage{subfigure}
\usepackage{rotating}

\shorttitle{On the 3D structure of the mass, metallicity, and SFR}
\shortauthors{Lara-L\'opez, L\'opez-S\'anchez & Hopkins}

\begin{document}

\title{On the 3D structure of the mass, metallicity, and SFR space for SF galaxies}

\author{Maritza A. Lara-L\'{o}pez} \email{mlopez@aao.gov.au}
\affil{Australian Astronomical Observatory, PO Box 915, North Ryde, NSW 1670, Australia}

\author{\'Angel R. L\'{o}pez-S\'{a}nchez}
\affil{Australian Astronomical Observatory, PO Box 915, North Ryde, NSW 1670, Australia, Department of Physics and Astronomy, Macquarie University, NSW 2109, Australia}

\author{Andrew M. Hopkins}
\affil{Australian Astronomical Observatory, PO Box 915, North Ryde, NSW 1670, Australia}

%


\begin{abstract}

We demonstrate that the space formed by the star-formation rate (SFR), gas-phase metallicity ($Z$), and stellar mass
(\Ms), can be reduced to a plane, as first proposed by Lara-L\'opez et al. We study three different approaches to find
the best representation of this 3D space, using a principal component analysis, a regression fit, and binning of the data.
The PCA shows that this 3D space can be adequately represented in only 2 dimensions, i.e., a plane. We find that the
plane that minimises the $\chi^2$ for all variables, and hence provides the best representation of the data,
corresponds to a regression fit to the stellar mass as a function of SFR and $Z$, \Ms=$f$(Z,SFR). We find
that the distribution resulting from the median values in bins for our data gives the highest $\chi^2$.
We also show that the empirical calibrations to the oxygen abundance used to derive the Fundamental Metallicity
Relation (Nagao et al.) have important limitations, which contribute to the apparent inconsistencies. The main
problem is that these empirical calibrations do not consider the ionization degree of the gas. Furthermore,
the use of the $N2$ index to estimate oxygen abundances cannot be applied for \abox~$\gtrsim8.8$ because
of the saturation of the [\ion{N}{ii}]\,$\lambda$6584 line in the high-metallicity regime.
Finally we provide an update of the Fundamental Plane derived by Lara-L\'opez et al.
\end{abstract}

\newpage

\section{Introduction}

Stellar mass (\Ms), metallicity ($Z$), and star-formation rate (SFR) 
are key galaxy properties.
\Ms\ reflects the amount of gas locked up in stars over a galaxy's history.
SFR indicates the current rate at which gas is being converted into stars. 
$Z$ reflects the gas reprocessed by stars over the course of stellar evolution, and any exchange of gas between the galaxy and the environment. 
The relationships between these three properties are fundamental in understanding galaxy evolution. Indeed, models of galaxy formation within the $\Lambda$-CDM scenario already include chemical hydrodynamic simulations 
\citep[e.g.,][]{DeLucia04,Tissera05,DeRossi06,Dave07,Martinez10}.

In the last few years it has been found that \Ms, $Z$, and SFR are strongly interrelated. Analyzing galaxy measurements from the Sloan Digital Sky Survey (SDSS), 
\citet{Ellison08} found that the mass-metallicity (\Ms$-Z$) relation for star-forming (SF) galaxies depends on the SFR.
Subsequently,  \citet{Lara10a}  reported the existence of a Fundamental Plane (FP) between these three parameters. These authors confirmed that the \Ms$-Z$ and \Ms$-$SFR relations are just particular cases of a more general relationship.  \citet{Lara10a} fitted a plane and derived an expression for  the stellar mass as a function of the gas metallicity and SFR (the Fundamental Plane, FP).
In a parallel and independent study, using the same SDSS data, but different Z and SFR estimations, \citet{Mannucci10} found a similar fundamental relationship, but instead expressed  $Z$ as a combination of  \Ms\ and SFR with a substantially different quantitative relationship. They refer to this correlation as the Fundamental Metallicity Relation (FMR).
In a recent study, \citet{Yates12} used models and SDSS data to analyze the dependences of different combinations between SFR, $Z$ and \Ms. They found qualitative differences in the  dependencies of those variables depending on the choice of approach in measuring the metallicity and SFR.


A  fundamental requirement in all these analyses is obtaining a reliable estimation of the galaxy metallicity.
The most robust method to derive the metallicity in SF galaxies is via the estimate of metal abundances and abundance ratios, in particular through the determination
of the gas-phase oxygen abundance. 
This is typically achieved through  the analysis of emission-line spectra of \HII\ regions within the galaxies. A proper determination of the oxygen abundance relies on the detection of the [\ion{O}{iii}]\,$\lambda$4363 auroral line \citep[the \Te\ method, e.g.,][]{LSE09},
but this emission line is usually not observed because of its faintness. Consequently, it is common to invoke the so-called strong-line methods. These techniques assume that the oxygen abundance of an \HII\ region can be derived using only a few bright emission lines. Empirical calibrations based on photoionization models, however, systematically over-predict by 0.2-0.6~dex the oxygen abundances derived using the \Te\ method and those calibrations which are based on it \citep[see][]{Yin+07,KE08,Bresolin09,LSE10b,Moustakas+10, LSDK+12}.  However, the absolute metallicity scale is still uncertain, temperature fluctuations and gradients can render the \Te\ method incorrect by up to 0.4 dex \citep{Peimbert07}.



Here we explore three different approaches to the representation of the three-dimensional
distribution of \Ms, SFR, and $Z$ for galaxies. We detail our sample selection in \S\,\ref{SampleSelection}
and review some issues with metallicity estimators in \S\,\ref{Metissue}. The analysis is presented in \S\,\ref{FPsection},
and we explore the implications for relationships between SFR and $Z$ in \S\,\ref{Z-SFR}. We present
a discussion of the outcome of our analysis, and summarise our results, in \S\,\ref{conc}.

\section{Sample selection}\label{SampleSelection}

We use data from the Sloan Digital Sky Survey Data Release 7 \citep[SDSS--DR7,][]{Abaza09,Adelman07}, using the emission-line analysis performed by the MPA-JHU group\footnote{http://www.mpa-garching.mpg.de/SDSS}. From the SDSS-DR7 database, we selected galaxies in the ``main galaxy sample" \citep{Strauss02}, with apparent Petrosian $r$ magnitude of \mbox{$14.5<m_r<17.77$} and in the redshift range $0.04<z<0.33$. The lower limit ensures that at least 20$\%$ of the galaxy light will be inside the $3''$ of the SDSS fiber, which is the minimum required to avoid aperture effects \citep{Kewley05}. To ensure reliable metallicities, we  imposed a minimum signal-to-noise ratio, SNR~$> 8$, for each of the most prominent lines \Ha, \Hb, \NII\,$\lambda$6584, [\ion{O}{iii}]\,$\lambda$5007, and {[O\,{\sc ii}]}\,$\lambda$3727.
The specific SNR threshold chosen turns out not to be critical to the results, and this is demonstrated in more detail in Section 4.4. The impact of SNR choice, and the
selection of lines it is imposed upon, is explored in detail by \citet[][]{Foster12}, who show that for SNRs
between about 3 and 8, the resulting mass-metallicity relation does not change substantially. For higher
SNR thresholds, though, the weaker (often lower SNR) lines may be excluded, leading to a reduced
sensitivity to the high-metallicity population.

We  selected SF galaxies following the criteria of \citet{Kauf03} on the \citet[][]{Baldwin81} diagram, \mbox{$\log$({[O\,{\sc iii}]}\,$\lambda$5007 / \Hb)} vs. \mbox{$\log$(\NII\,$\lambda$6584 / \Ha).}
The above criteria give us a sample of 45\,475 galaxies.

We use gas-phase metallicities, total SFRs, and stellar masses derived by the MPA-JHU group.
Metallicities were estimated through a Bayesian approach based on simultaneous fits to all the most prominent lines according to \citet[][hereafter, T04]{Tremonti04}, while total SFRs were obtained from \citet{Brinchmann04}.
Aperture effects are again an important issue to consider in the estimate of SFR \citep{Brinchmann04}. \citet{Yates12}
shows that if SFRs are calculated without making an aperture correction, nearby galaxies will have their SFRs
underestimated, and this leads to a reduced spread in the SFR distribution for low mass, low redshift galaxies. This
will have an impact on any exploration of the SFR dependence of the \Ms$-Z$ relation.
Both \citet{Tremonti04} and \citet{Brinchmann04} make use of the population synthesis and photoionization codes given by \citet{Charlot01}. Stellar masses were estimated from fits to the photometry as described in \citet{Kauf03}.

\section{Metallicity estimate issues}\label{Metissue}

Since the calculation of metallicity is a particularly challenging process, the most accurate and
reliable measurements available should be used whenever possible. Approximating metallicity
estimates through simple parameterisations can be a valuable tool when only limited information
is at hand, but such approximations have significant limitations and uncertainties.

\begin{figure*}[t]
 \centering
\includegraphics[scale=0.50]{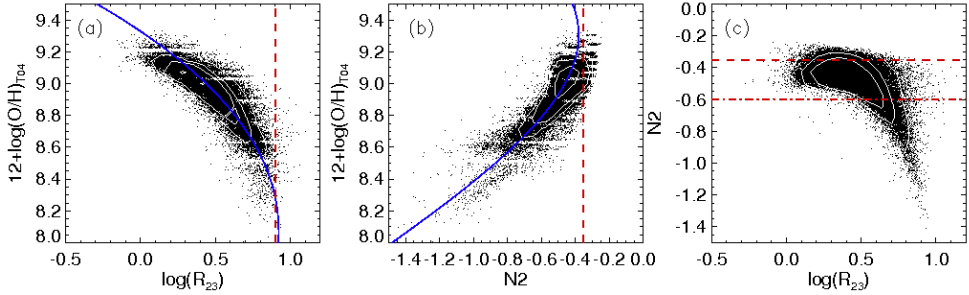}
\caption{Panels $a$ and $b$  compare the $\log(R_{23})$ and $N2$ parameters with the T04 metallicities, respectively. The blue line gives the N06 calibrations, while the red dashed line indicates the limit imposed by  \citet{Mannucci10} to their cubic fits. Panel $c$ compares the $R_{23}$ and $N2$ indexes. The red dashed line indicates the limit of panel $b$, while the dot-dashed line shows the limit in which the $N2$ method should be avoided ($N2$$>-0.6$).}
\label{MetNagao}
\end{figure*}

In this section we analyze several metallicity indicators, including the empirical calibrations of \citet[][hereafter, N06]{Nagao06}, which are cubic fits between the $R_{23}\equiv$ ([O\,{\sc ii}]\,$\lambda$3727+ [O\,{\sc iii}]\,$\lambda\lambda$4959,5007)/\Hb\  and the \mbox{$N2\equiv \log$(\NII\, $\lambda$6584\,/ \Ha)} parameters and the T04-derived metallicity. We analyze the following methods:

\begin{description}

\item[] (\textit{i}) $\;$ The $R_{23}$ parameter and the calibration of N06, which can be used only when log($R_{23}$) $< 0.90$, see Fig. \ref{MetNagao}a.
\vspace{-2mm}\item[] (\textit{ii}) $\;$ The $N2$ parameter and the calibration of N06, which can be used only when $N2 < −0.35$, see Fig. \ref{MetNagao}b.
\vspace{-2mm}\item[] (\textit{iii})$\;$ The mean value of the $N2$ and $R_{23}$ (hereafter mean($N2$,$\;$$R_{23}$)) where the two metallicity values differ by less than 0.25 dex, as used in \citet[][]{Mannucci10}.
\vspace{-2mm}\item[] (\textit{iv})$\;$The  T04 bayesian metallicities, which take into account an ionization parameter.
\vspace{-2mm}\item[] (\textit{v})$\;$The  \NII$\lambda$6563 / \OII$\lambda$3727 ratio (hereafter \NII/\OII ), which has been shown to be a very reliable metallicity indicator \citep[e.g.][]{KD02, KE08}. For this ratio, we use the \citet[][]{KD02} calibration, and recalibrate those metallicities to the T04 system using 12+log(O/H)=-0.7329+1.0841 [ 12+log(O/H)$_{\rm KD02}$].

\end{description}


It is well known  \citep[e.g.][]{Baldwin81,McGaugh91,KD02,PT05,LSE10b,LS+IC10+11} that the degree of ionization of the gas plays a fundamental role in deriving a reliable estimation of the oxygen abundance.
The most common empirical calibrations based on either a direct determination of the \Te\ \citep{PT05,PVT10} or  photoionization models  \citep{McGaugh91, KD02,KK04,Tremonti04} do consider an ionization parameter. Without such information, the uncertainty of the derived oxygen abundances may be as large as $0.25-0.40$~dex \citep{LSE10b}. 

We test how sensitive the N06 calibrations are to the ionization parameter ($U$). According to \citet{Dors11}, the [S\,{\sc ii}] $\lambda\lambda6717,6731$ and \Ha \ lines give an accurate indicator which is almost independent of reddening.  We estimate the ionization parameter for our entire sample using the prescription of \citet{Dors11}:

{ \small \begin{equation}\label{IonPot}
 \rm log (U)=-1.66 \ \rm log ([\rm SII] \lambda\lambda6717,6731 / {\rm H\alpha} ) -4.13
\end{equation}}

The ionization parameter represents the dimensionless ratio of the ionizing photon density to the electron density. A metallicity diagnostic that takes into account the ionization parameter would reduce the uncertainty in the derived metallicities and thus reduce the scatter against log(U). On the other hand, metallicity diagnostics based on empirical fits that do not take into account any ionization parameter will increase the uncertainty in metallicity and show a high dispersion against log(U).

\begin{figure*}
 \centering
\includegraphics[scale=0.62]{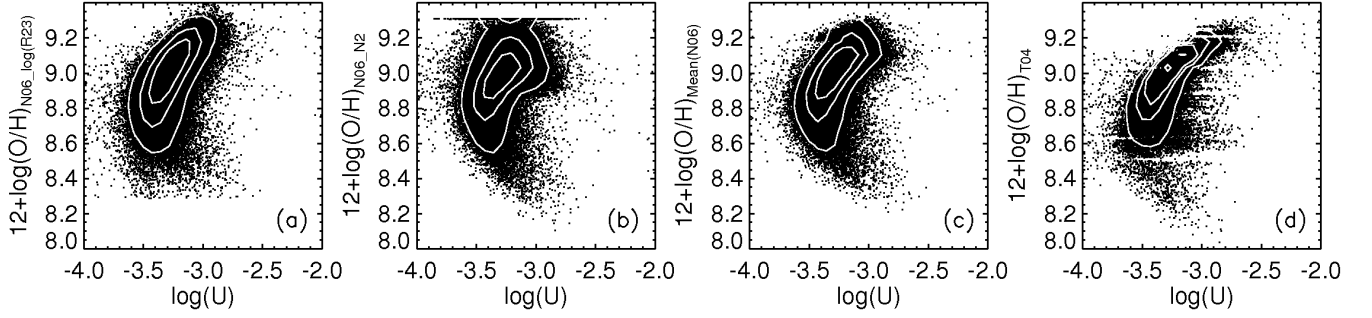}
\caption{Comparison between metallicity and the ionization parameter for the ($a$) $R_{23}$ method, ($b$) $N2$ parameter, ($c$) mean($N2$, $R_{23}$), and ($d$) Tremonti et al. (2004).}
\label{Ionization}
\end{figure*}

A comparison of the ionization parameter with the N06 metallicities using the $N2$, $R_{23}$, mean($N2$, $R_{23}$), and the T04 metallicities are shown in Fig. \ref{Ionization}. It can be appreciated that the N06 metallicities show a high dispersion against log(U). Also, it is clear from Fig. \ref{Ionization}b that metallicities higher than $\sim$ 8.8 show the highest scatter. 
We also plot the T04 metallicities for comparison in Fig. \ref{Ionization}d, which show a very tight correlation because those take into account the ionization parameter. Considering just the narrow metallicity range of 9.1-9.15, for example, we find that the $\sigma$ in the scatter of log(U) is 0.1, 0.13, 0.14, and 0.07 dex for $R_{23}$, $N2$, mean($N2$, $R_{23}$), and the T04 metallicities. The highest dispersion of log(U) in this range of metallicities is given by mean($N2$, $R_{23}$). It is clear that although the $N2$ metallicities have being averaged with the $R_{23}$ metallicities, the resultant is still strongly affected by the saturation and sensitive to the ionization parameter of the $N2$ index.

We highlight that although the N06 calibrations are based on the T04 metallicities, which do consider an ionization parameter, this does not mean that the N06 calibrations are corrected for ionization. The most robust approach is clearly to estimate the ionization parameter for every galaxy, to provide the most accurate metallicy estimation. When global fits are used to estimate metallicities, the uncertainty in metallicity can be as high as 0.15 dex, as shown in Fig. \ref{MoreComparison}. As a consequence, the simplified fits of N06 should be avoided when sufficient emission lines are available to make a more reliable and direct estimate of the metallicity \citep{LSE10b}.


\begin{figure*}
 \centering
\includegraphics[scale=0.62]{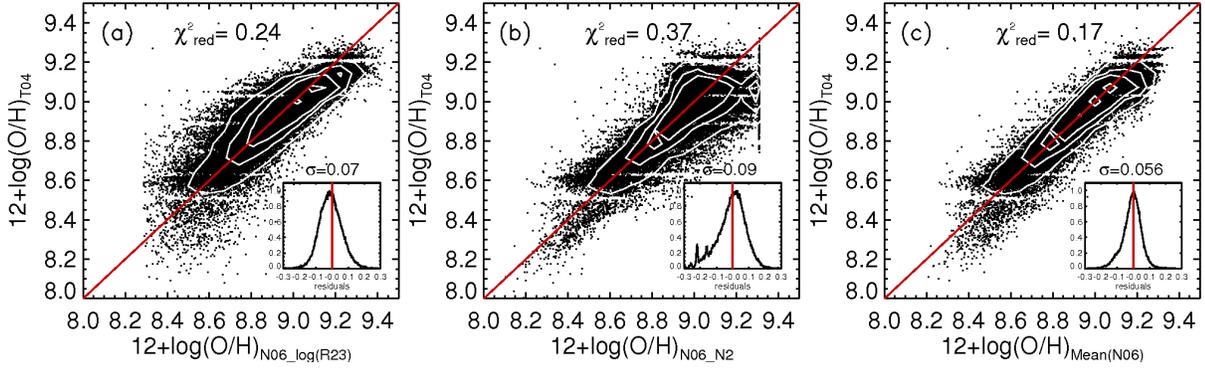}
\caption{Panels ($a$) and ($b$) show the metallicity obtained  by the N06 calibrations using the $\log(R_{23})$ and $N2$ indexes vs. the T04 metallicities, respectively. Panel ($c$) shows the mean($N2$, $R_{23}$). The inset shows the histogram of the residuals.}
\label{N2Comparison}
\end{figure*}

\begin{figure*}
 \centering
\includegraphics[scale=0.62]{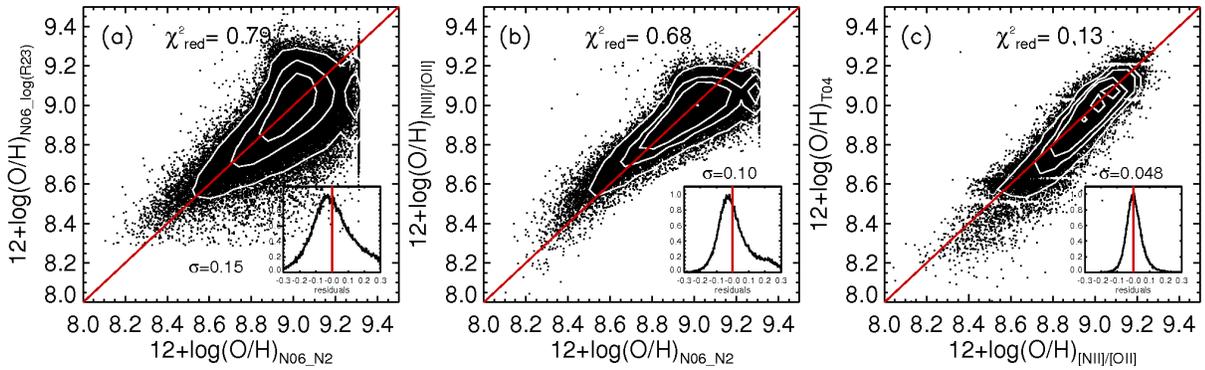}
\caption{Panels ($a$) and ($b$)  compare the $N2$ metallicities with the $R_{23}$ method and the \NII /\OII ratio, respectively. Panel ($c$) compare metallicities using the T04 and \NII /\OII ratio. The inset shows the histogram of the residuals.}
\label{MoreComparison}
\end{figure*}

Furthermore, the use of  the $N2$ parameter to derive metallicities is not valid in the high metallicity regime. This can be appreciated in Fig. \ref{MetNagao}b, c, and \ref{N2Comparison}b , which shows the relationship between the T04-derived metallicity  and the $N2$ parameter and metallicity, respectively. It is clear that the $N2$ index saturates for \abox~$\gtrsim8.8$.

To further support this, and to caution authors against using calibrations in a regime in which they are not valid, we performe a statistical analysis comparing all the metallicity indicators described in this section. Throughout  we use the reduced chi-squared as a measure of goodness of fit, defined by the following equation:

\begin{equation}\label{ChiSq}
{\chi}_{red}^2= \frac{1}{\nu} \sum   \frac {(O_i-E_i)^2}{\sigma^2}
\end{equation}
where  $O$ and $E$ are the observed data and model estimate, respectively. The value $\nu$
is the number of degrees of freedom given by $\nu=N-n-1$, where $N$ is the number of observations and $n$ is the number of fitted parameters. $\sigma^2$  is the variance of the observations defined by:

\begin{equation}\label{Variance}
{\sigma}^2= \frac{1}{N} \sum {(O_i-\mu)^2}
\end{equation} where $\mu$ is the mean of the observed data.

In Figure \ref{MoreComparison} we compare the $N2$ metallicities with the $R_{23}$ and \NII / \OII \  calibrations. Regardless of the method used, It can be clearly seen that the $N2$ parameter always saturates for metallicities higher than $\sim$8.8 (Figs. \ref{N2Comparison}b, \ref{MoreComparison}a,  \ref{MoreComparison}b), resulting in a very high $\chi^2_{red}$ and $\sigma$. As a sanity check, we compare the T04 metalicities with those obtained using the \NII / \OII \  ratio (Fig. \ref{MoreComparison}c), obtaining a tight correlation with the lowest  $\chi^2_{red}$ and $\sigma$ of all the comparisons. Table \ref{MetComparison} shows a summary of the $\chi^2_{red}$ and $\sigma$ for all our metallicity comparisons.

\begin{table}[t]
\begin{center}
\begin{tabular}{c||cc}
\hline
\hline
Compared methods  & $\chi^2_{red}$ & $\sigma$ \\\hline
\NII / \OII vs. T04	&  0.13	&	0.048\\\
Mean($N2$, $R_{23}$) vs. T04		&	0.17&	0.056	\\\
$R_{23}$ vs. T04	&	0.24	&	0.07\\\
$N2$ vs. T04	&	0.37	&	0.09\\\
$N2$ vs. \NII / \OII &	0.68	&	0.10\\\
$N2$ vs. $R_{23}$	&	0.79	&	0.15\\\hline
\noalign{\smallskip}
\end{tabular}
\normalsize
\rm
\end{center}
\caption{Summary of $\chi^2_{red}$ and $\sigma$ of the residuals for several metallicity methods.}
\label{MetComparison}
\end{table}

\begin{center}
\begin{figure}[t]
 \centering
\includegraphics[scale=0.28]{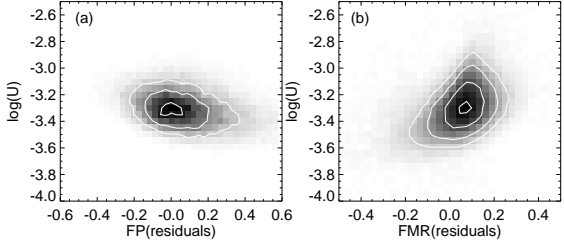}
\caption{The ionization parameter is shown against the residuals from (a) the Fundamental Plane (Eq. \ref{FPEq} of this paper), and (b) the Fundamental Metallicity Relation of \citet{Mannucci10}.}
\label{FPyFMPIon}
\end{figure}
\end{center}

This  agrees with the result found by \citet{Yin+07}, who concluded that empirical calibrations involving the $N2$ parameter are only valid for galaxies with \abox$<$8.5, when comparing with \Te-based abundances; this limit should be 8.7--8.9 when $Z$ has been derived using photoionization models, as in our case \citep{LSDK+12}.

Finally, we show the ionization parameter as a function of residuals for the FP and FMR (Fig. \ref{FPyFMPIon}). In this representation we are using the N06 calibrations (mean($N2$, $R_{23}$) ) and  \citet[][]{Mannucci10} Eq. 4 for the FMR. For the FP we are using the MPA-JHU data described above, and Eq. \ref{FPEq} described below in Sect. 4.2. It is clear from Fig. \ref{FPyFMPIon} that the ionization parameter is relatively flat with the residuals from the FP, but increases proportionally with the residuals from the FMR. This suggests that scatter around the FMR is likely a consequence of the N06 calibrations neglecting the ionization parameter.


The use of the $N2$ parameter in a metallicity regime for which it is not valid should therefore be avoided. Even when this method is being averaged with another \citep[e.g.][]{Mannucci10, Nagao06}, this can still drastically affect the resulting dependencies between SFR, \Ms, and Z. For example, Figure 1 of \citet{Yates12} shows the SFR dependence on the \Ms$-Z$ relation using the N06 and T04 metallicities. It can be
appreciated from that figure that the dependence on SFR disappears in the high metallicity regime when
the N06 metallicities are used. This lack of dependence can be explained by the saturation of
the $N2$ parameter in the high metallicity regime. On the other hand, when the T04 metallicities
are used, this dependence is observed over the full metallicity range. To further support this observation, \citet[][]{Lara13} analyzed the same dependence using several combinations of
metallicity \citep[e.g.,][]{PP04, Tremonti04, KD02} and SFR \citep[e.g.][]{Hopkins03,Brinchmann04} indicators,
obtaining for all the possible combinations a strong dependence of SFR in the \Ms$-Z$ relation for the
full metallicity range.

\section{A 3D analysis of the \Ms, Z, and SFR space}\label{FPsection}


Here we aim to identify the most compact representation of the data distribution in the 3D space
of \Ms, $Z$ and SFR. We examine three methodologies: $(i)$ fitting a plane to the 3D distribution using
PCA, $(ii$) fitting a plane through regression \citep{Lara10a}, and $(iii)$ binning in SFR and \Ms\ to obtain
the median metallicity in each bin \citep{Mannucci10}.

As we want to perform a self-consistent comparison of the results, and because of the metallicity issues described
in \S\,\ref{Metissue}, we do not use the \citet{Mannucci10} method to derive $Z$, in the third approach
mentioned above. Rather we test all the  approaches self-consistently using the MPA-JHU measurements
detailed \S\,\ref{SampleSelection}.

%
%

\begin{figure*}
\begin{center}
\begin{tabular}{cc}
\includegraphics[scale=0.28]{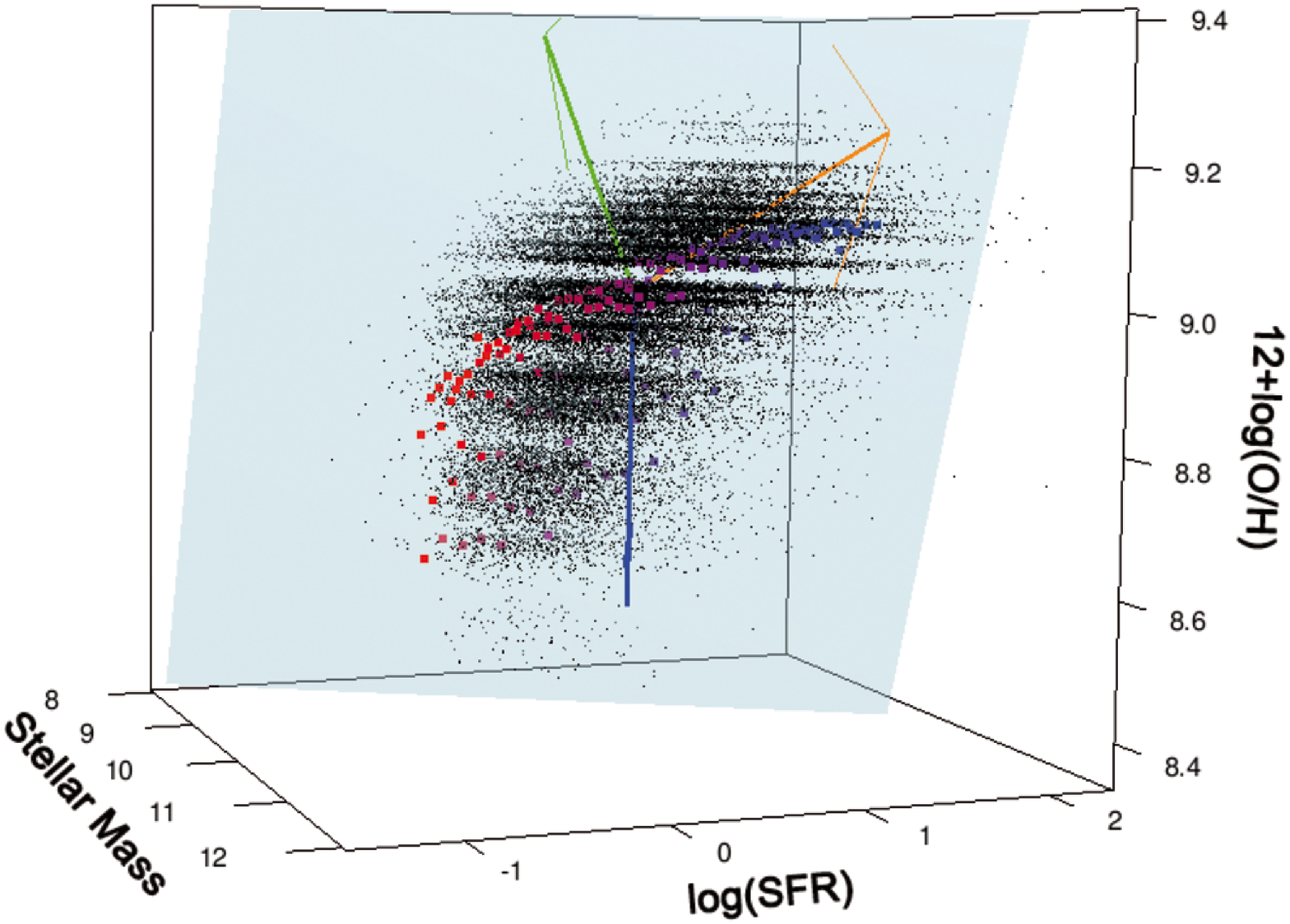}&
\includegraphics[scale=0.28]{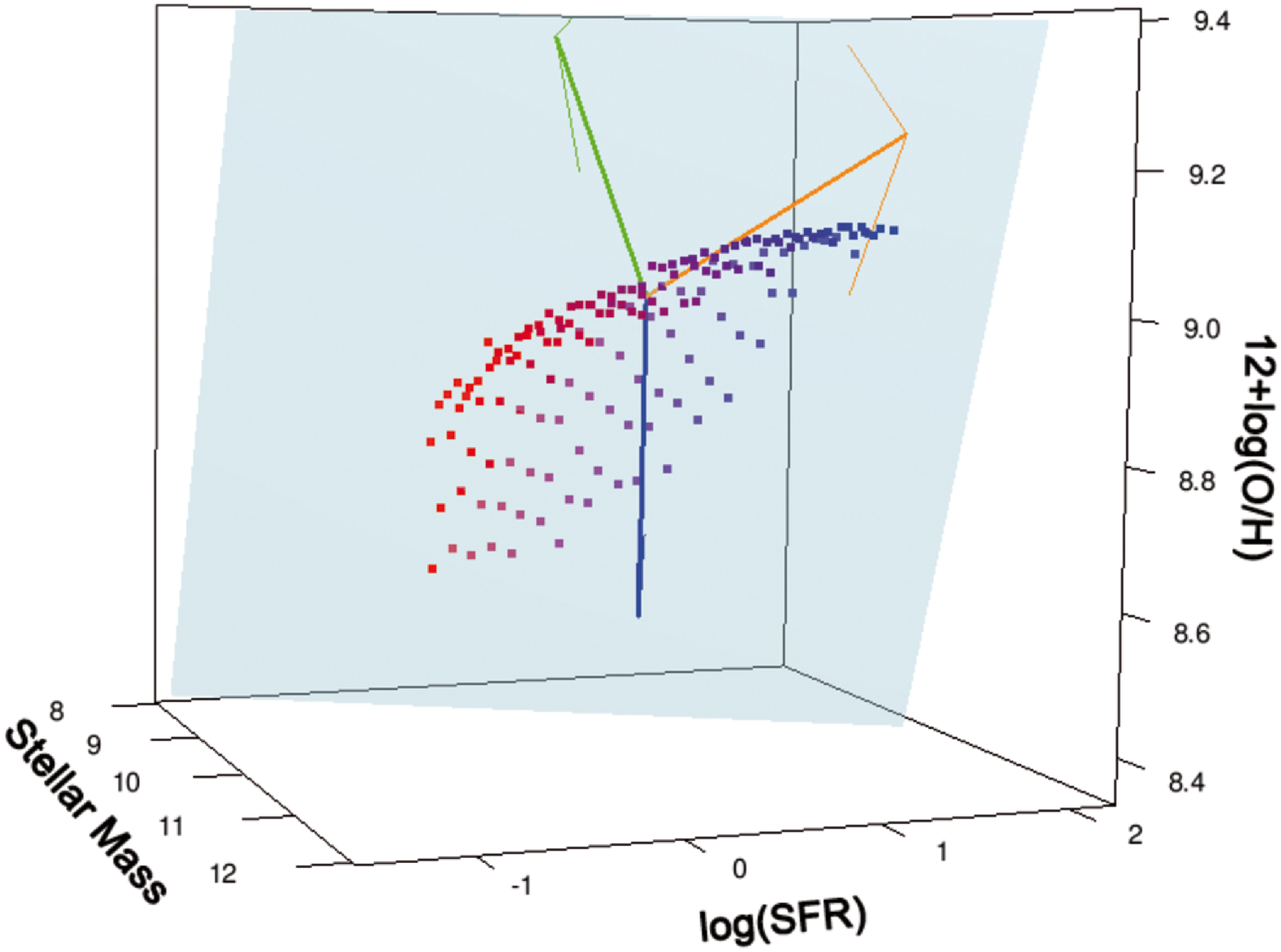} \\
\includegraphics[scale=0.28]{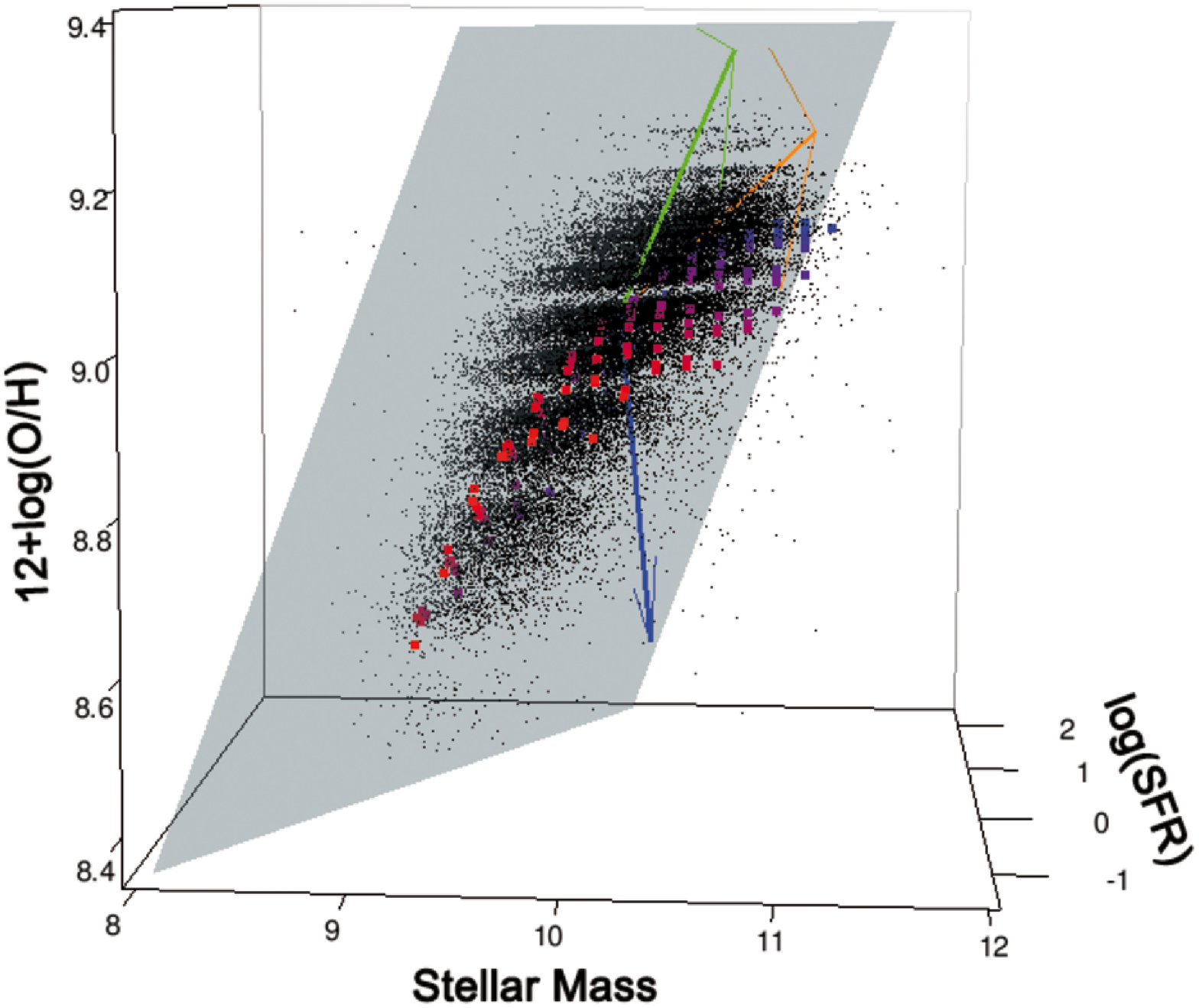}&
\includegraphics[scale=0.28]{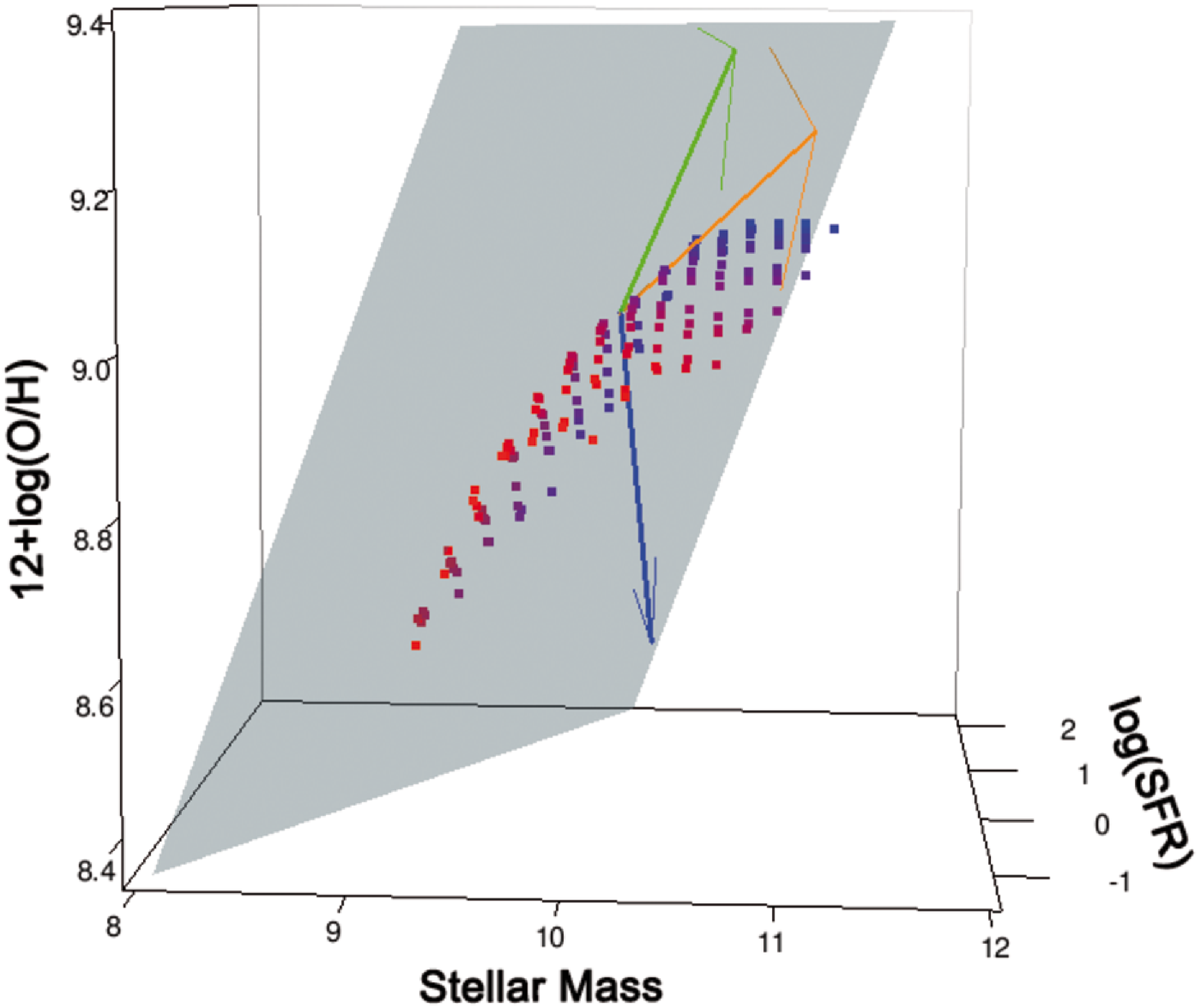}\\
\includegraphics[scale=0.28]{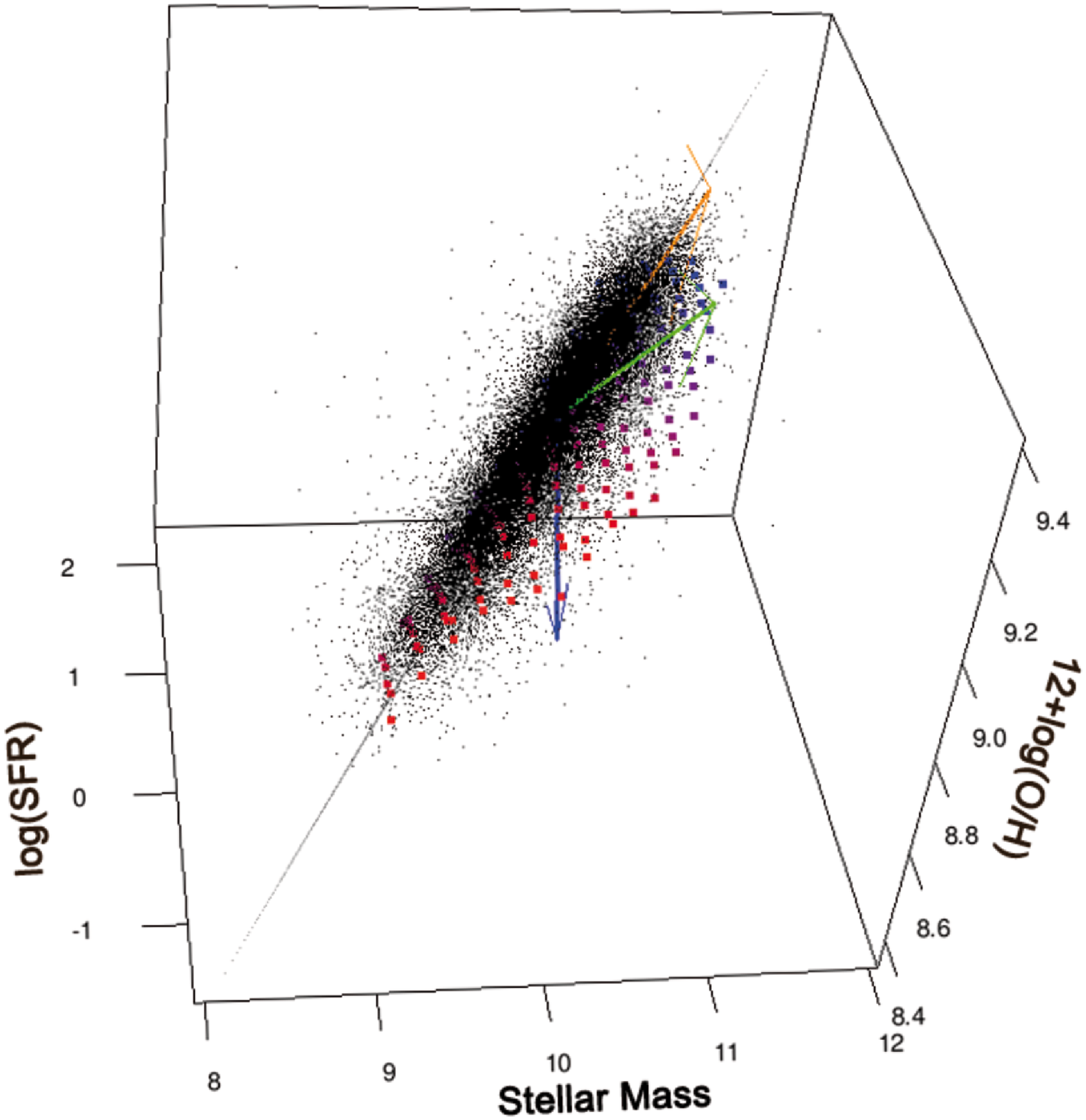}&
\includegraphics[scale=0.28]{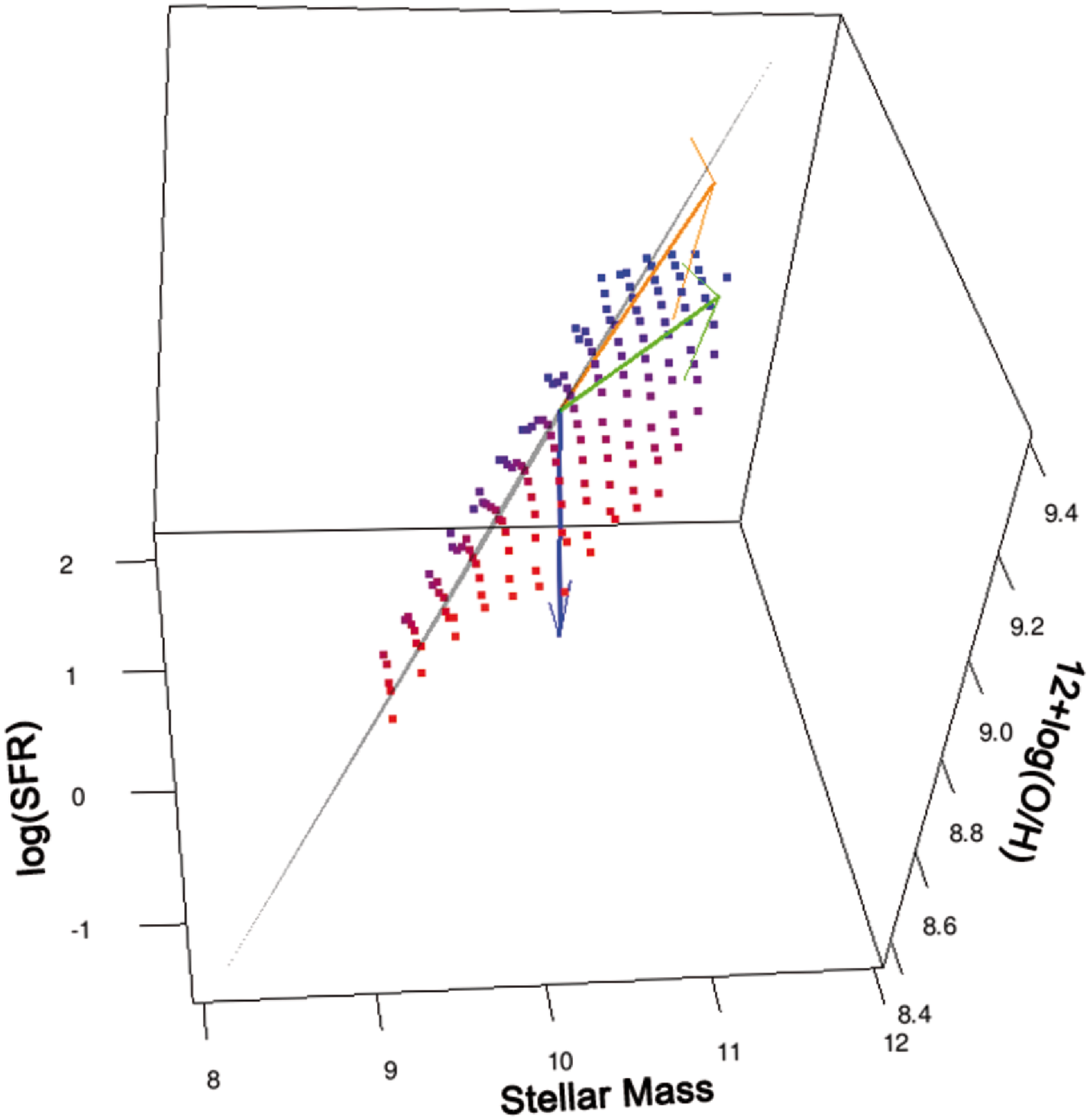}\\
\end{tabular}
\end{center}
\caption{Different orientations of the 3D space formed by \Ms, SFR, and $Z$. The metallicity is always kept on the vertical axis, the cube is just moved rightwards, then tilted forward slightly, in going from the top to the bottom panels. The left panels show our newly derived FP (shaded), while the colored square points show the median metallicity taken in bins of SFR and \Ms\  \citep[as for the FMR,][]{Mannucci10}. Square points are color-coded from low (red) to high (blue) SFR.
Black data points are the full sample (\S\,\ref{SampleSelection}).  The vectors show the first PCA component in yellow, the second in green, and the third in blue. The right panels show the same orientation and information as the left panels but omitting the underlying sample of SDSS galaxies. Upper panels show a face-on view of the FP, middle panels show an angle close to the \Ms$-Z$ relation, while bottom panels show the FP at the lowest dispersion face.}
\label{Cubos3D}
\end{figure*}

\subsection{PCA analysis}

We  performed a principal component analysis (PCA) to identify the underlying dimensionality of the three observables. PCA is a mathematical procedure that converts a set of observations of possibly correlated variables into a set of uncorrelated variables called principal components. One of the goals of PCA is to reveal hidden structure in a dataset, as well as to reduce the dimensionality of the data. A high correlation between variables is an indicator of high redundancy in the data, while the most important and independent variables are those that account for the largest variance.
Since high correlation is a mark of high redundancy, the principal components should have low or even zero correlation between them \citep{Shlens09}.


We find that the first two principal components  account for  86$\%$ and 12$\%$ of the variance, which indicates that 98$\%$ of our data can be explained in a 2 dimensional space. As a result of the PCA procedure we obtain 3 eigenvectors,
referred to here as comp1, comp2, comp3, which are expressed as the vector of coefficients of the three 
parameters, $x$=\Ms, $y$=12+log(O/H), and $z$=log(SFR). The first principal component indicates the direction
of the highest variance, and is given by comp1$=(0.7140, 0.1679, 0.6796)$, while the second and third components account for the highest possible variance in orthogonal directions: comp2$=(0.5952, 0.3654, -0.7156)$, comp3$=(0.3686, -0.9155, 0.1609)$, see Fig.~\ref{Cubos3D}.  It is important to note here that one of the weak points of PCA is that it relies
on the covariance matrix, which is less robust against outliers than other methods. We return to this point in
the discussion below.

The plane obtained through PCA is given by:

{ \small \begin{equation}\label{PCAplane}
\alpha \ [{\rm{M_{\star}}}/{\rm M_{\odot}}] + \beta \ [12+\rm log(O/H)] + \gamma \ [log(SFR)] = \delta
\end{equation}} where  $\alpha=0.3686,\ \beta=-0.9155,\ \gamma=-0.1609, \ \delta=-4.5578$

\begin{figure*}
\begin{center}
\includegraphics[scale=0.31]{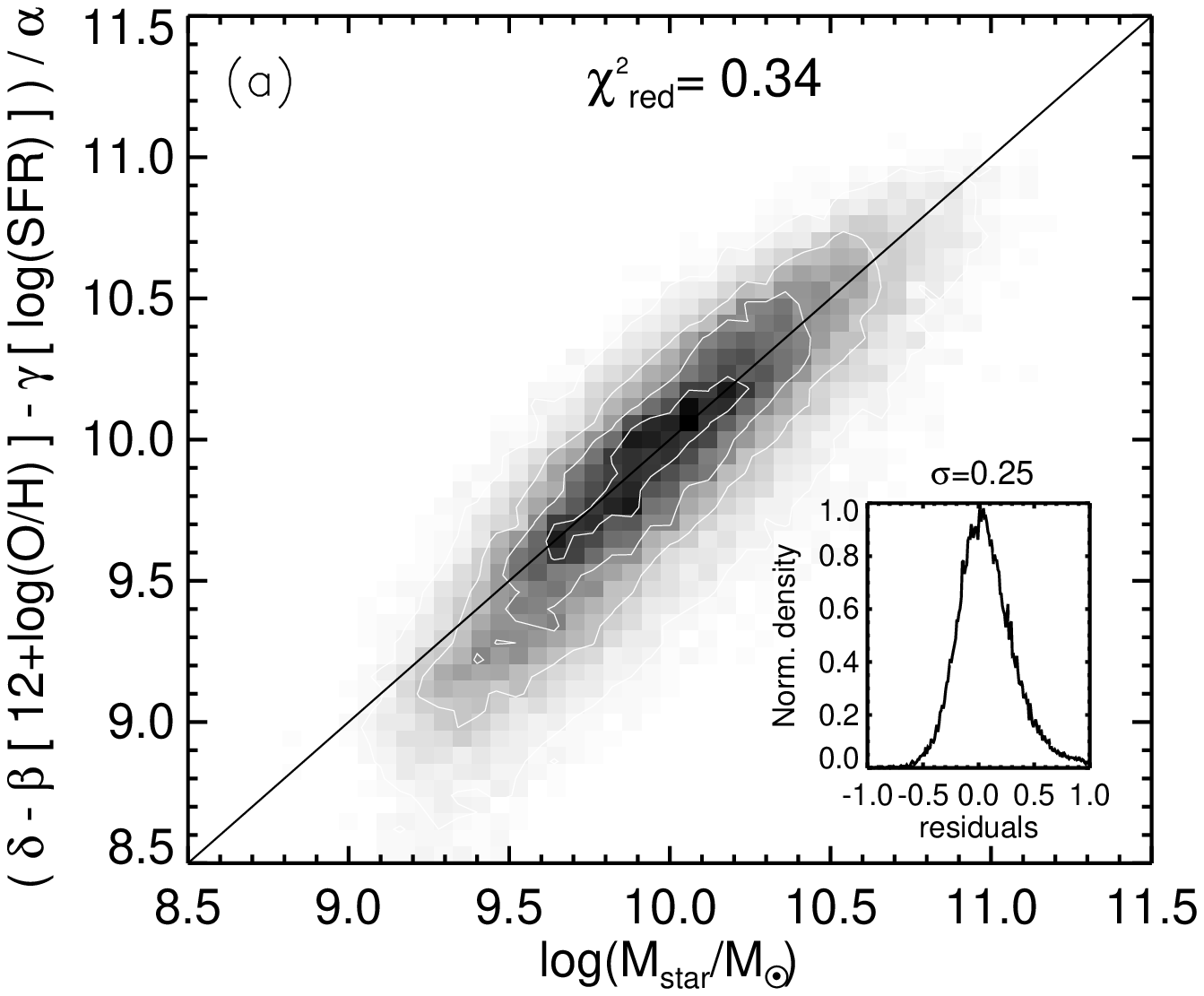}
\includegraphics[scale=0.32]{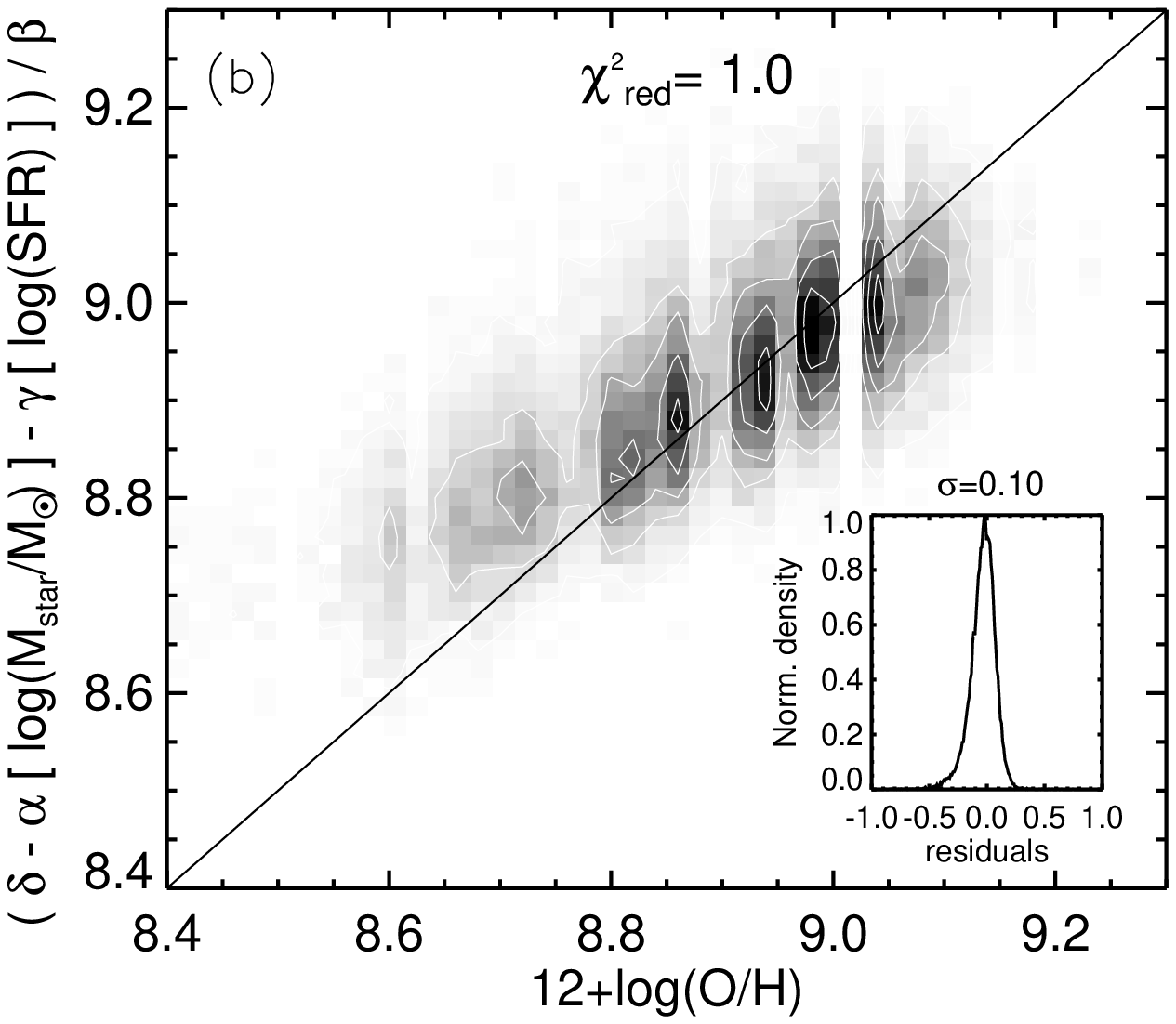}
\includegraphics[scale=0.32]{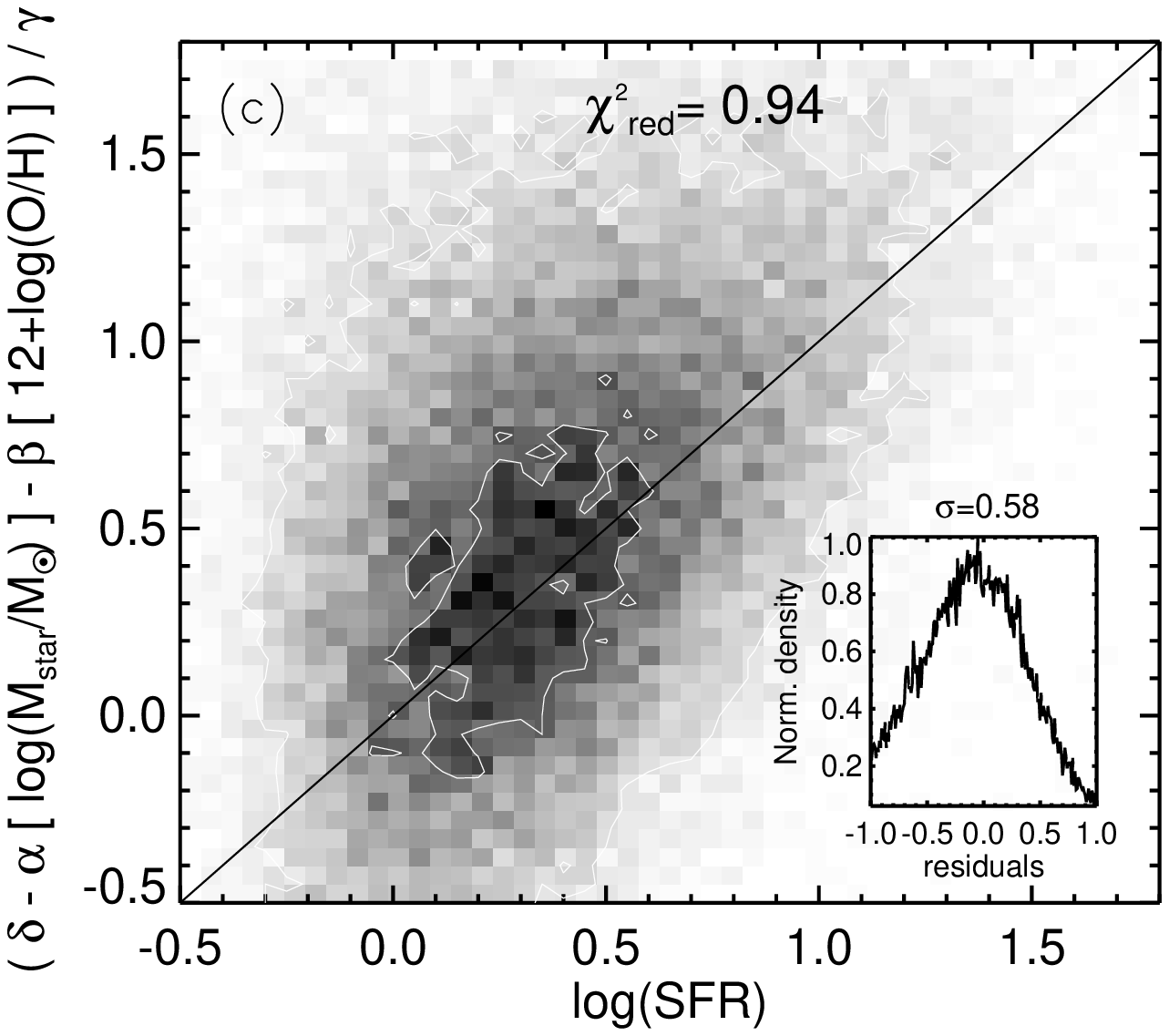}
\end{center}
\caption{PCA plane. From left to right, \Ms, metallicity and SFR estimated through PCA. The solid line shows the one to one relation, and the inset shows the histogram of the residuals.}
\label{PCA}
\end{figure*}

Solving Eq.~\ref{PCAplane} in turn for each of \Ms, $Z$ and SFR, as a function of the other two,
we obtain the relations shown in Fig. \ref{PCA}. Although PCA gives an acceptable result in
reproducing the \Ms\ with $\chi_{red}^2=0.34$, the metallicity and especially the SFR present a high $\chi_{red}^2$
of $1.0$ and $0.94$, respectively, (Fig.~\ref{PCA}). It can also be appreciated in Fig.~\ref{PCA}
that the fit provided by PCA is less effective at capturing the shape of the distribution, in particular at the
low mass and metallicity end. This is a consequence of the covariance matrix sensitivity to outlying data points.
This sensitivity effectively pushes the principal components to try to represent extremes in the data
that are not representative of the bulk of the measurements.


\begin{figure*}
\begin{center}
\includegraphics[scale=0.32]{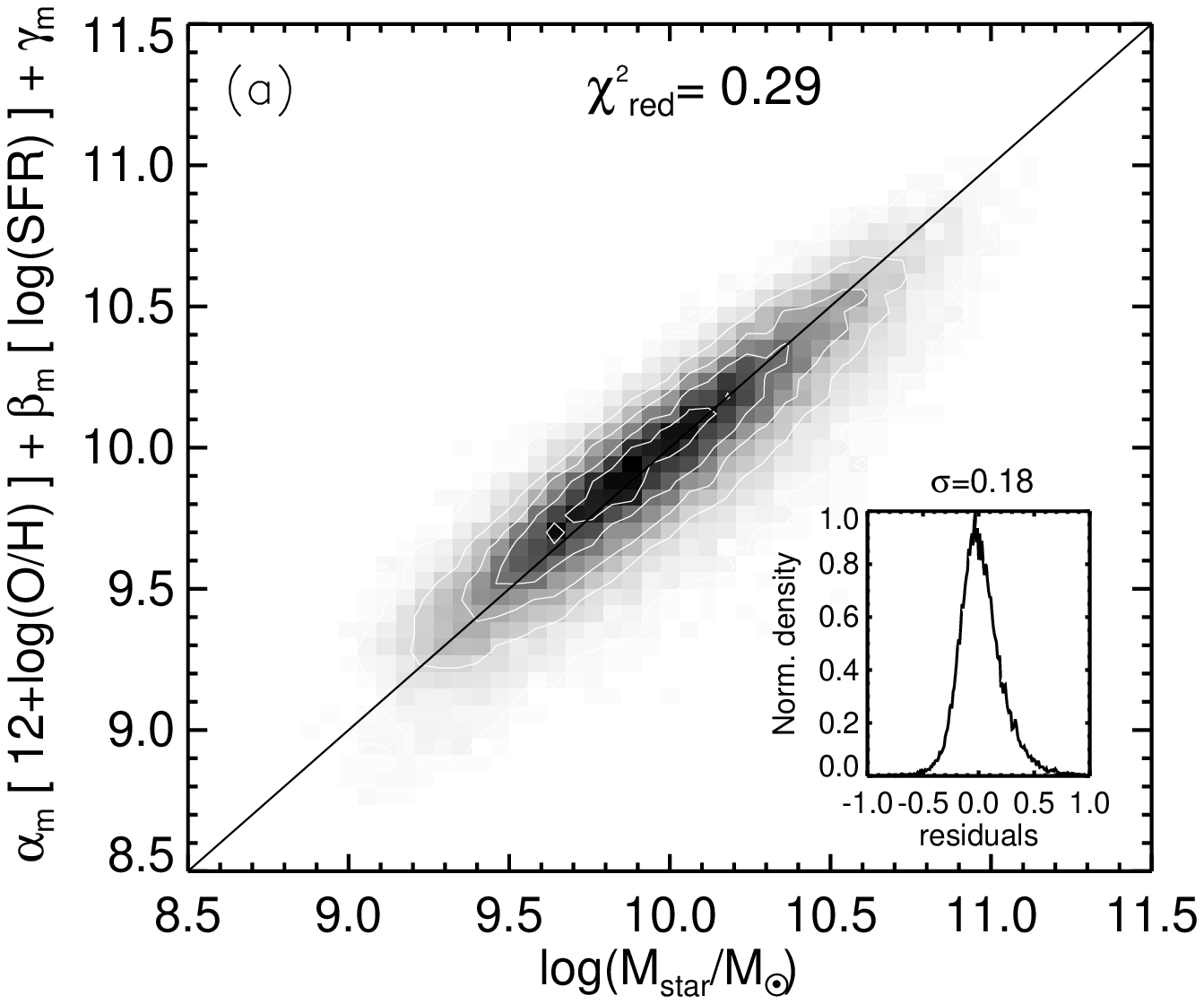}
\includegraphics[scale=0.32]{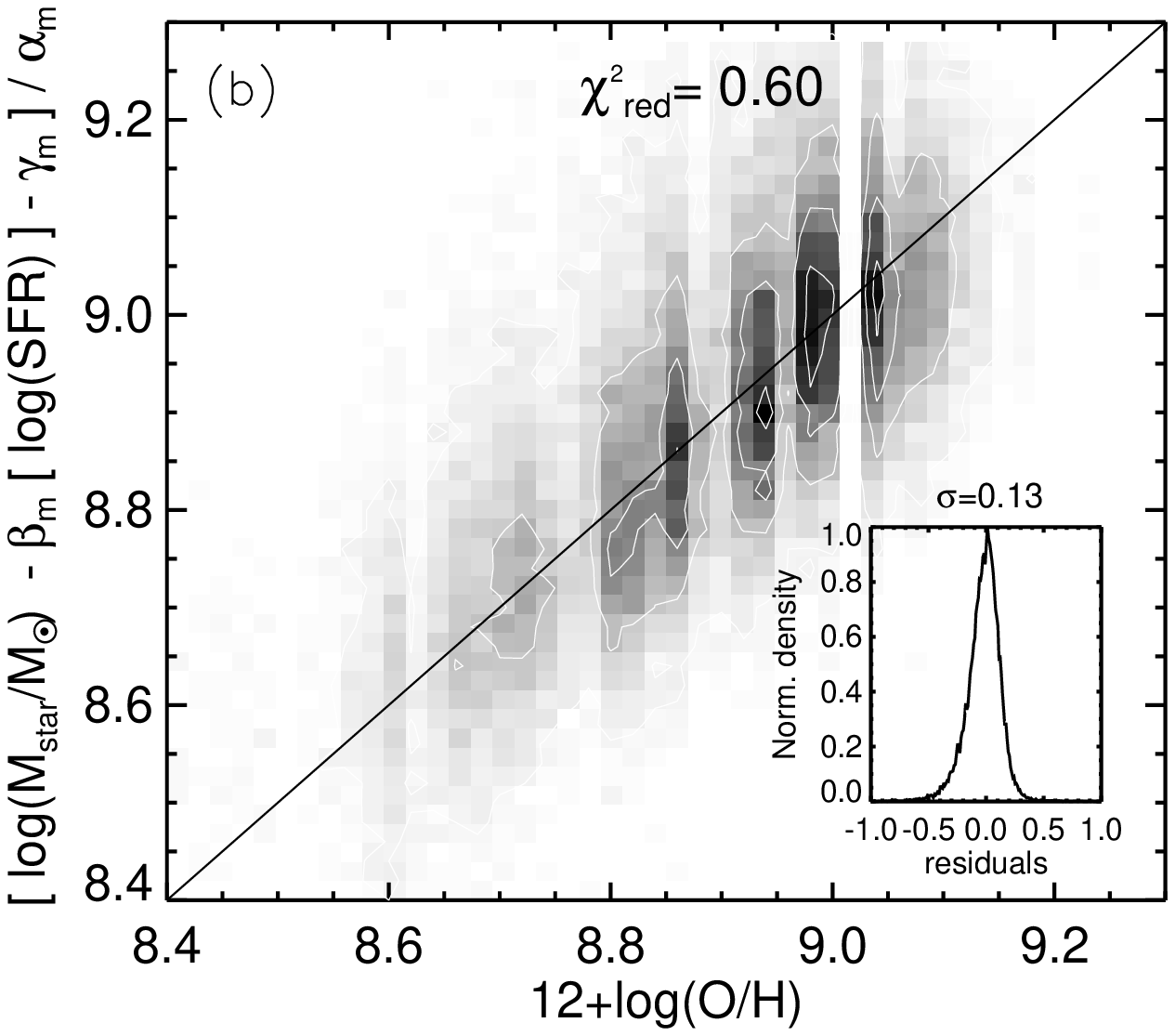}
\includegraphics[scale=0.32]{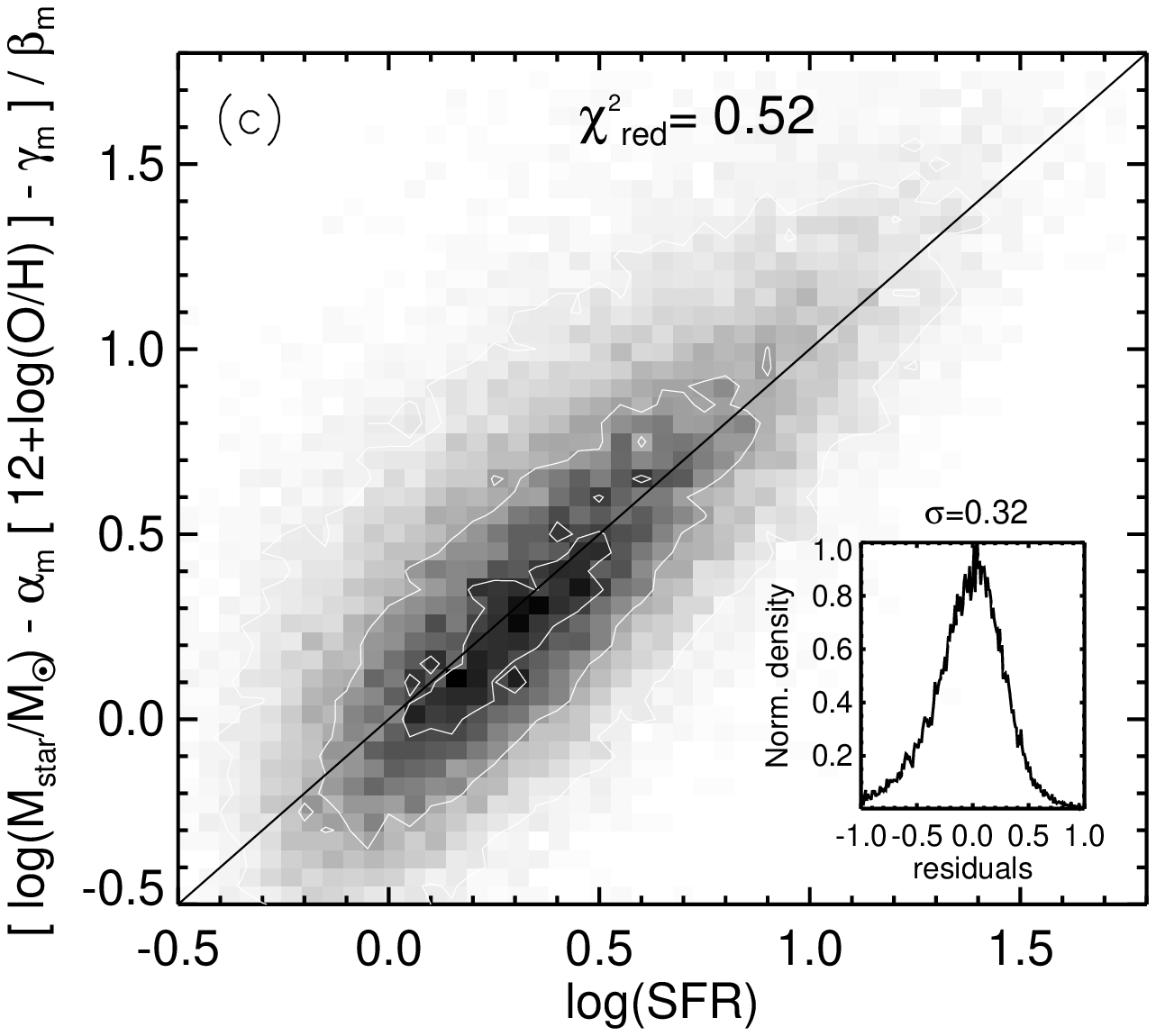}
\end{center}
\caption{Plane fitted to \Ms\ using regression,  \Ms=$f$(Z, SFR), the FP approach of \citet{Lara10a}.}
\label{FPEstMasa}
\end{figure*}

\begin{figure*}
\begin{center}
\includegraphics[scale=0.32]{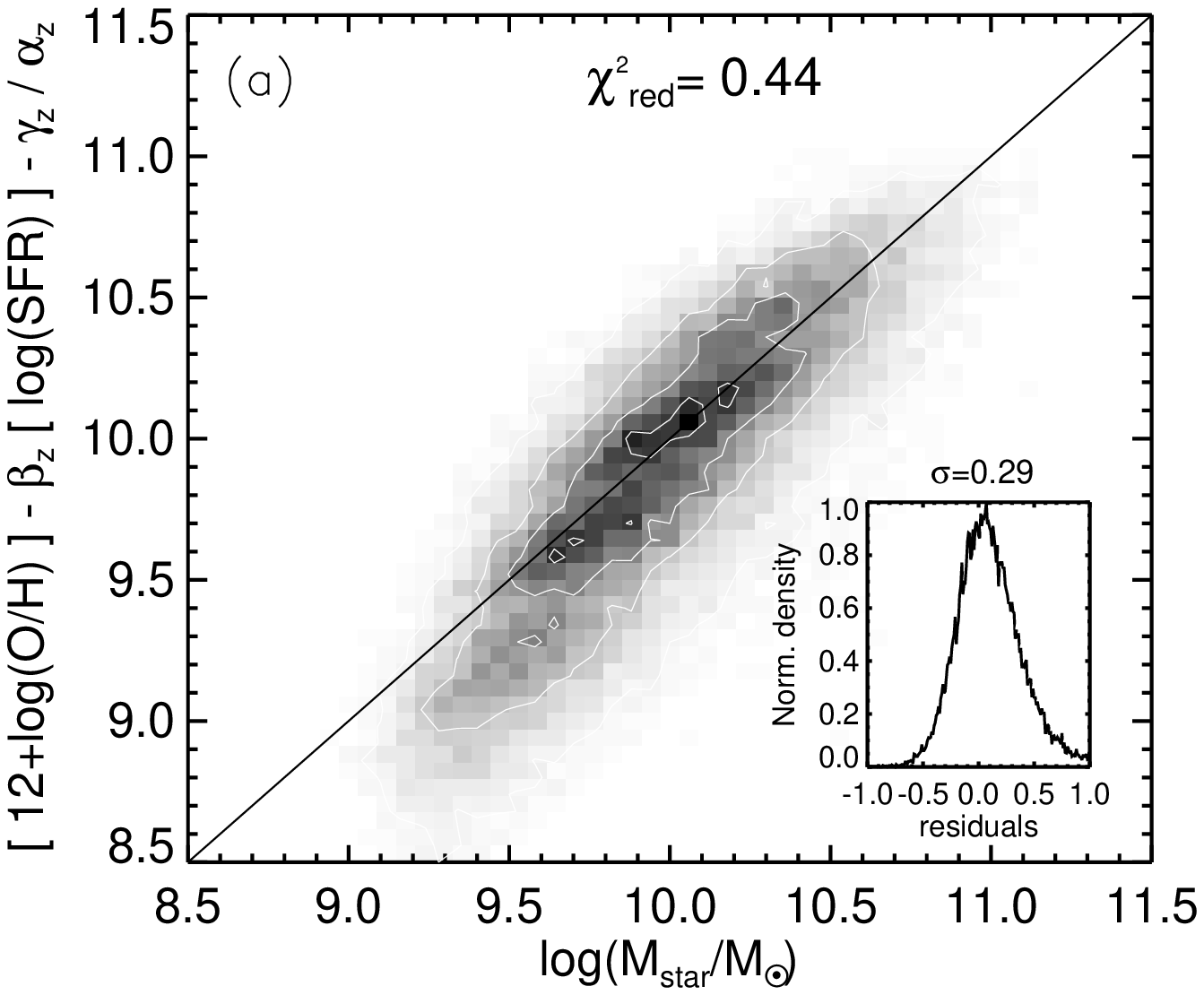}
\includegraphics[scale=0.32]{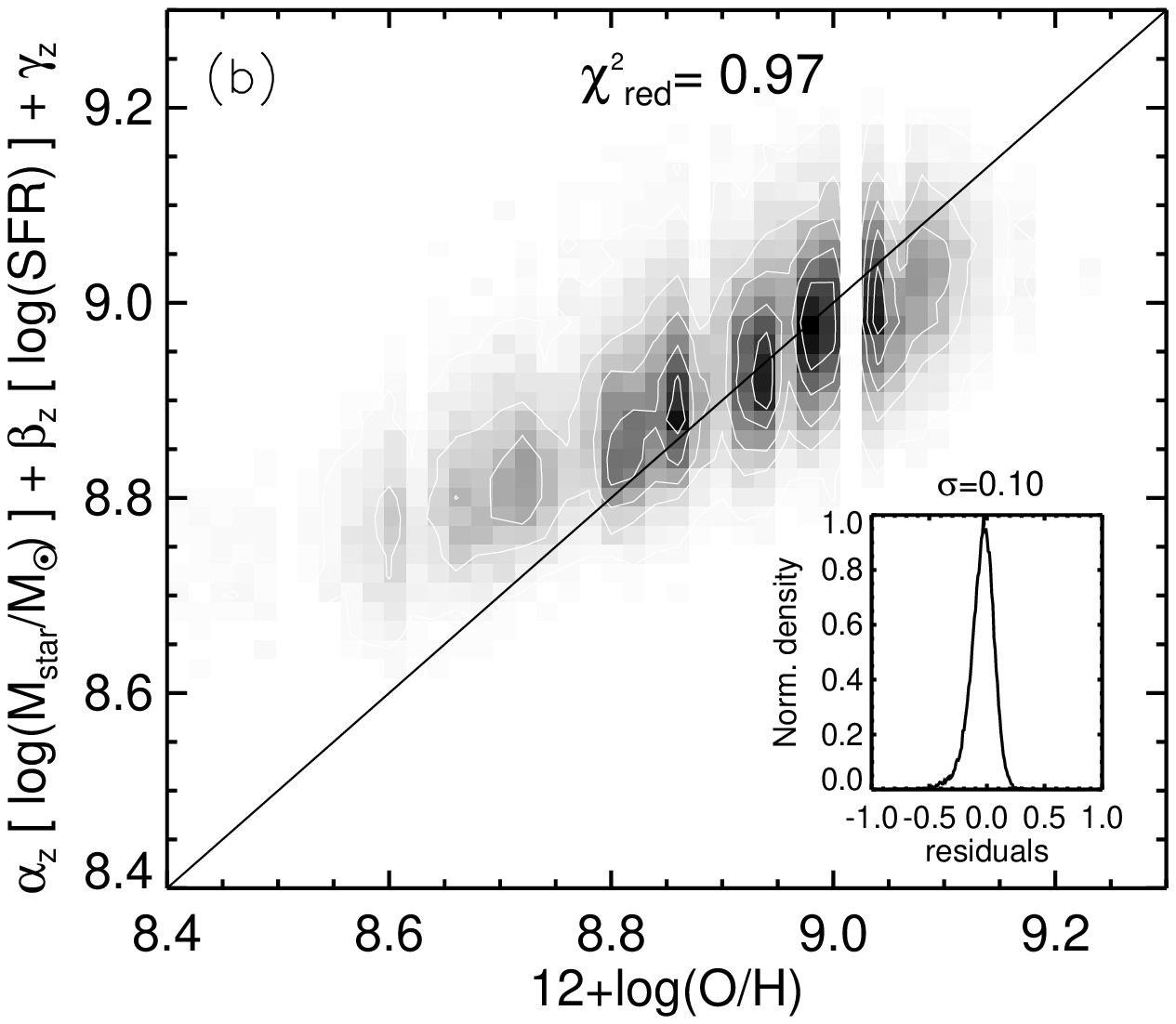}
\includegraphics[scale=0.32]{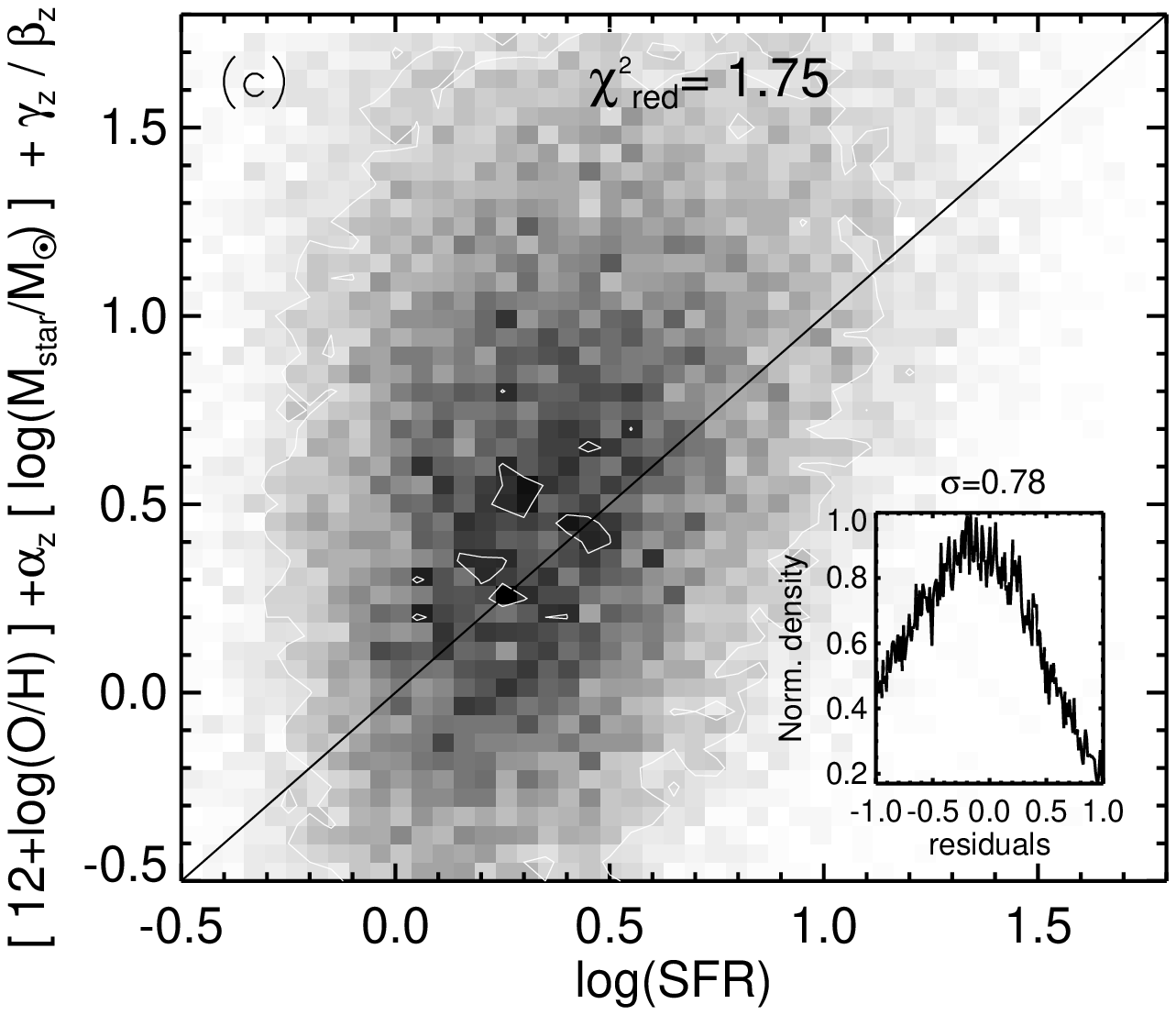}
\end{center}
\caption{Plane fitted to $Z$ using regression,  Z=$f$(\Ms, SFR).}
\label{FPEstZ}
\end{figure*}

\begin{figure*}
\begin{center}
\includegraphics[scale=0.32]{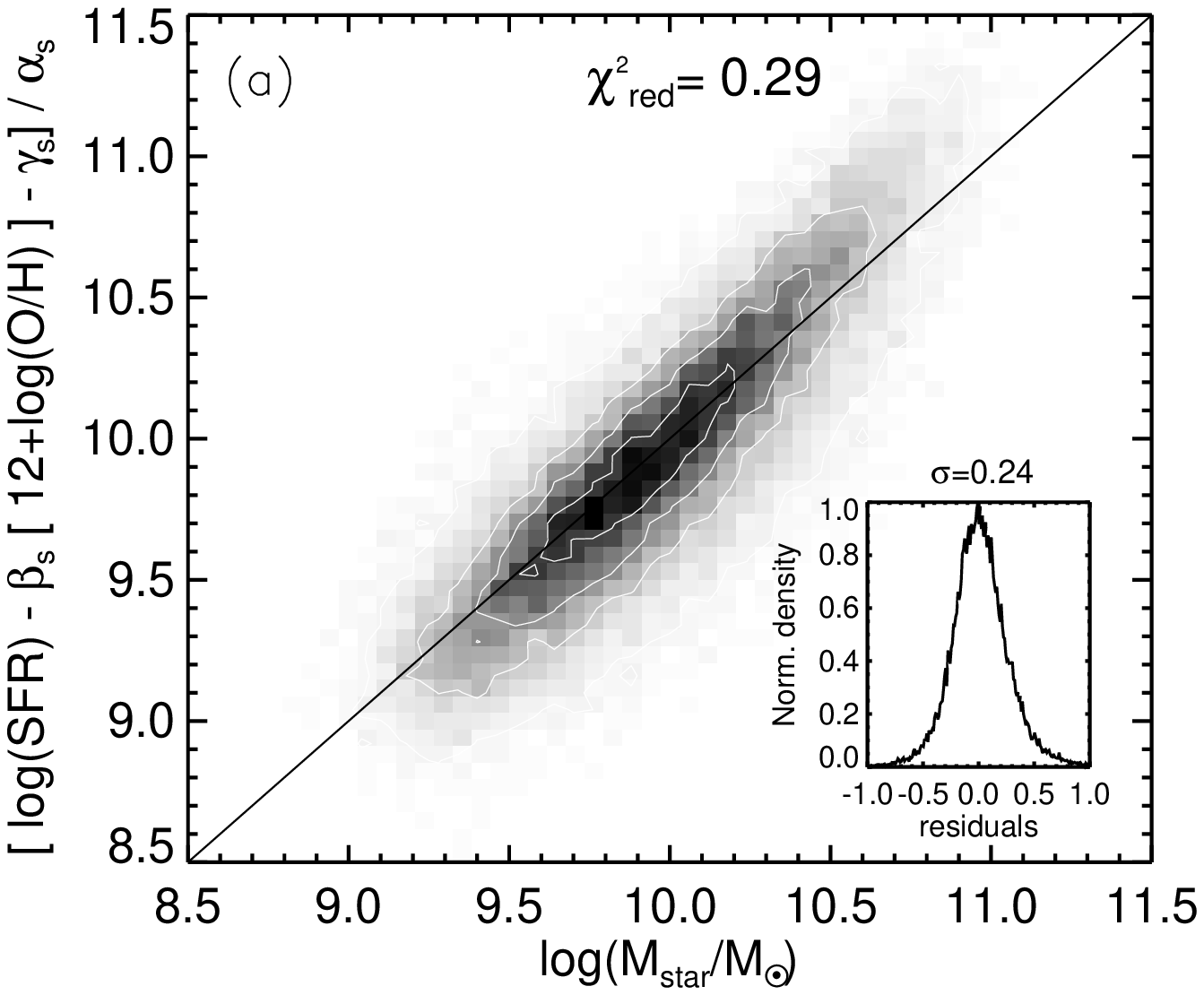}
\includegraphics[scale=0.32]{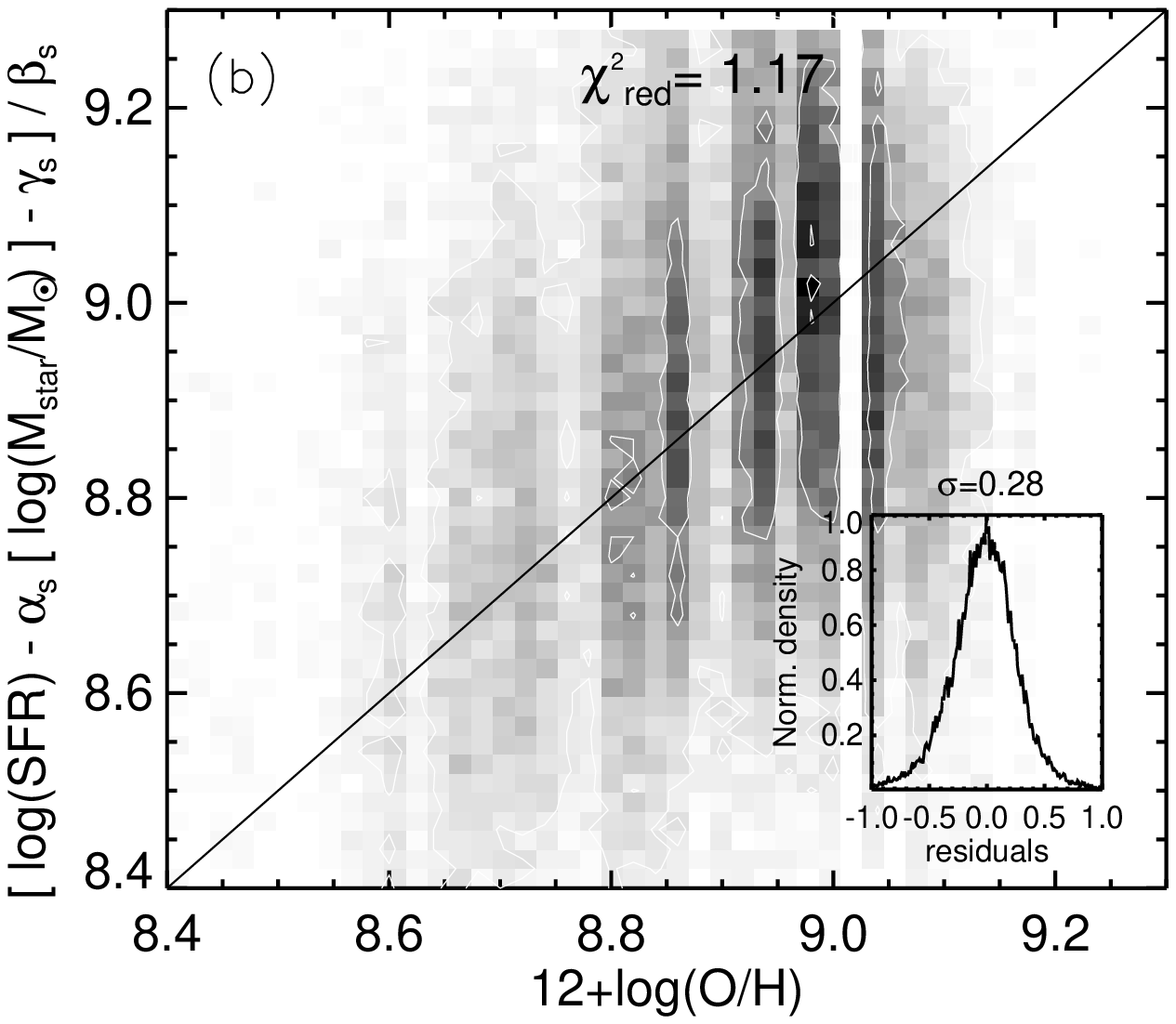}
\includegraphics[scale=0.32]{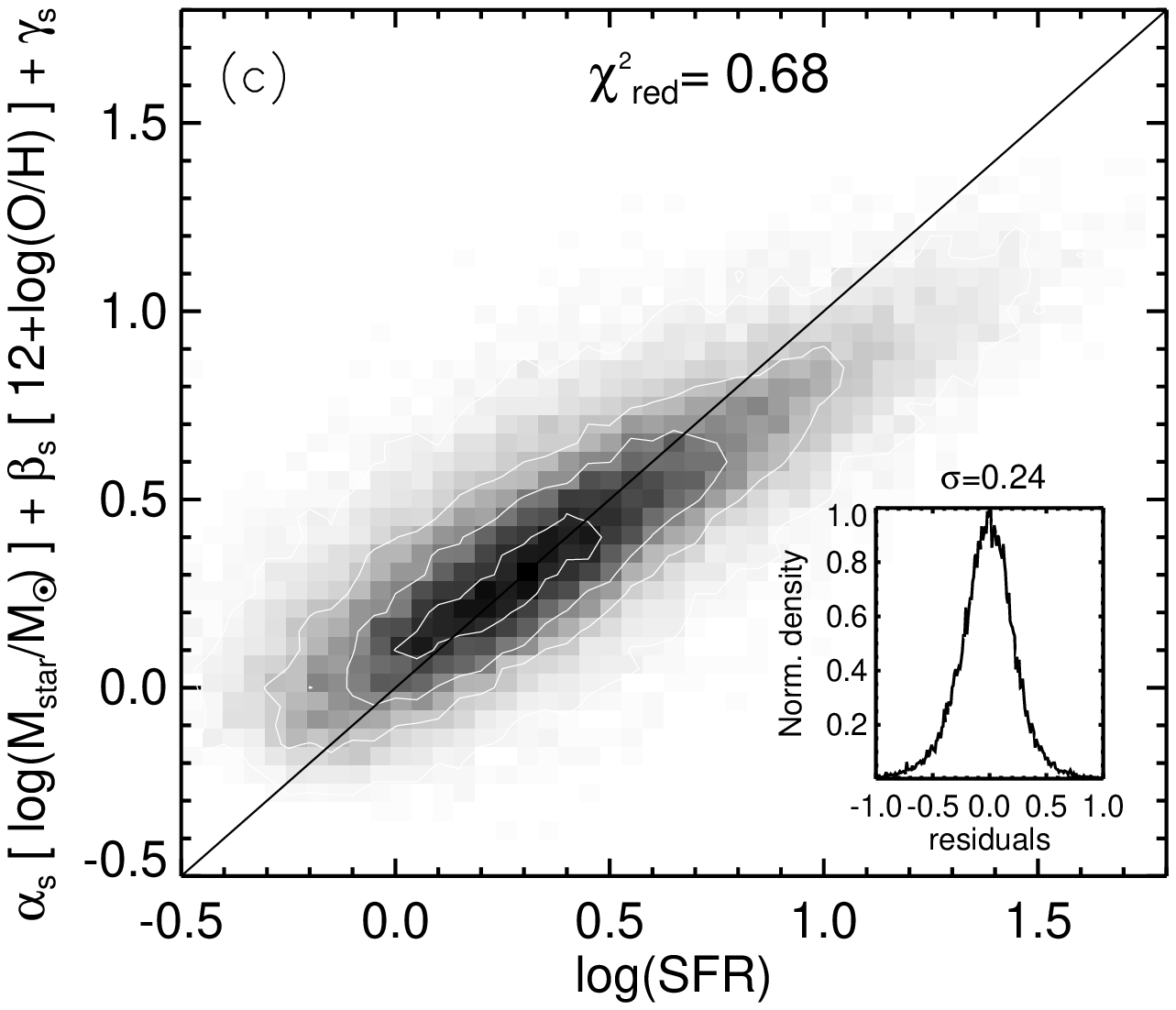}
\end{center}
\caption{Plane fitted to SFR using regression,  SFR=$f$(Z, \Ms).}
\label{FPEstSFR}
\end{figure*}

\begin{table*}
\begin{center}
\begin{tabular}{c||cc||cc||cc}
\hline
\hline
\multicolumn{1}{c} {}&\multicolumn{2}{c} {\Ms}&\multicolumn{2}{c}{Z}&\multicolumn{2}{c}{SFR}\\ \cline{1-7}
Fit  & $\chi^2_{red}$ & $\sigma$ & $\chi^2_{red}$ & $\sigma$ & $\chi^2_{red}$  & $\sigma$  \\ \hline
PCA	&	0.34		&	0.25	&	1.0	&	0.1	&	0.94	&	0.58	\\\
\Ms$=f(Z,{\rm SFR})$ (FP) 	&	0.29	&	0.18	&	 0.60	&	 0.13	&	0.52	&	0.32	\\\
$Z=f($\Ms$,{\rm SFR})$	&	0.44	&	0.29	&	0.97	&	0.1	&	1.75	&	 0.78	\\\
SFR$=f($\Ms$,Z)$ &	0.29	&	0.24	&	1.17	&	0.28	&	0.68	&	0.24	\\\
FMR	&		--	&	--	&	1.06 	&	0.12	&	--	&	-- \\\hline
\end{tabular}
\normalsize
\rm
\end{center}
\caption{Summary of $\chi^2_{red}$ and $\sigma$ of the residuals for \Ms, metallicity and SFR using
planes derived by PCA, regression, and the FMR.}
\label{TableChi}
\end{table*}

\subsection{Regression}\label{FPsection_reg}

We next explore regression to represent our 3D data distribution. Regression aims to explain one variable
in terms of the others, and uses robust methods that are less affected by outlying observations. Using regression
we first fit a plane to \Ms\ as a function of SFR and $Z$, obtaining:

{ \small \begin{equation}\label{FPEq}
{\rm log}({\rm{M_{\star}}}/{\rm M_{\odot}}) = \alpha_m \ [12+\rm log(O/H)] + \  \beta_m \ [log(SFR)] \ + \gamma_m \ 
\end{equation}} where  $\alpha_m=1.3824,\ \beta_m=0.5992,\ \gamma_m=-2.5729$.

The plane derived in this way was called the Fundamental Plane (FP) by \citet{Lara10a} and is represented in Fig.~\ref{Cubos3D} in blue. The difference between Eq.~\ref{FPEq} and that presented in \citet{Lara10a} is that here we are including galaxies at higher redshifts, up to $z \sim 0.35$. This improves the sampling of the high mass galaxy population, thus slightly changing
the orientation of the plane. A detailed discussion of redshift and mass completeness will be given in
Lara-L\'opez et al.\ (in preparation).

Although the FP (Eq.~\ref{FPEq}) is defined to minimize the variance in the estimate of \Ms, we also tested it to estimate the metallicity and SFR  of galaxies by rearranging Eq. \ref{FPEq} to solve for the other variables (Fig.~\ref{FPEstMasa}). Estimating $\chi_{red}^2$ as before, we see that the metallicity obtained through the FP gives
$\chi_{red}^2=0.60$. Fig.~\ref{FPEstMasa}b also demonstrates that the estimate of $Z$ follows a linear behaviour
over the full metallicity range. We also tested this FP to estimate the SFR of galaxies, shown in Fig.~\ref{FPEstMasa}c, obtaining $\chi_{red}^2=0.52$, which is an improvement compared to the $\chi_{red}^2=0.94$ obtained for
the SFR through PCA. Again, Fig.~\ref{FPEstMasa}c shows a more linear relationship through the full range
of SFR compared to PCA.

To test whether the choice of \Ms\ as the dependent variable in Eq.~\ref{FPEq} is the optimal approach,
we also analyzed the two other possible planes, fitting in turn to $Z$ and SFR (Fig.~\ref{FPEstZ} and
Fig.~\ref{FPEstSFR}).

Fitting a plane to $Z$ as a function of SFR and \Ms, we obtain:

{ \small \begin{equation}\label{FP_ZEq}
12+\rm log(O/H)=\alpha_z \ [{\rm{M_{\star}}}/{\rm M_{\odot}}] + \beta_z \ [log(SFR)] + \gamma_z 
\end{equation}} where  $\alpha_z=0.3504,\ \beta_z=-0.1289,\ \gamma_z=5.4882$, giving $\chi_{red}^2=0.97$ for the metallicity. To compare this plane with the others, we again also estimate \Ms\ and SFR by rearranging Eq.~\ref{FP_ZEq}. We find $\chi_{red}^2=0.44$ for \Ms, and $\chi_{red}^2=1.75$ for SFR (Fig.~\ref{FPEstZ}).

We also fit SFR as a function of $Z$ and \Ms\ to obtain:

{ \small \begin{equation}\label{FP_SFREq}
\rm log(SFR)=\alpha_s \ [{\rm{M_{\star}}}/{\rm M_{\odot}}] + \beta_s \ [12+\rm log(O/H)] + \gamma_s 
\end{equation}} where  $\alpha_s=0.9924,\ \beta_s=-0.8511,\ \gamma_s=-1.9167$, and giving $\chi_{red}^2=0.68$
for the SFR. Again, we estimate \Ms\ and $Z$ by rearranging Eq.~\ref{FP_SFREq}.
This gives $\chi_{red}^2=0.29$ for \Ms, and $\chi_{red}^2=1.17$ for $Z$.

\subsection{Binning data}

Following \citet{Mannucci10}, we generated a grid of 0.11~dex in $\log$(SFR), and 0.15~dex in \Ms\ and estimated the median metallicity in every square of the grid. The resulting values are shown in Fig.~\ref{Cubos3D}. This figure only shows median values for those bins containing at least 50 galaxies. It can be seen that, despite the different metallicity estimator, we can reproduce the shape of the surface obtained by \citet{Mannucci10}. By showing the underlying data used in deriving this surface (left panels of Fig.~\ref{Cubos3D}), it becomes clear that the curvature in this surface is not representative of the actual data distribution.

\begin{figure}[t]
\begin{center}
\includegraphics[scale=0.5]{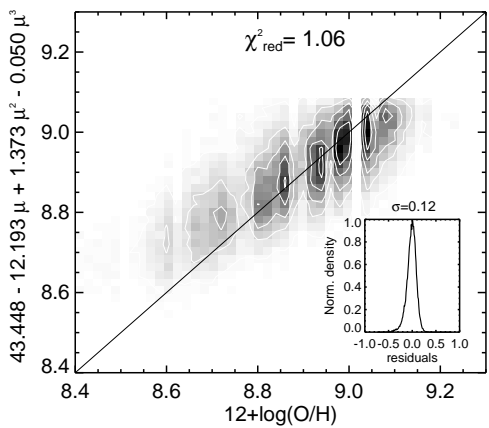}
\end{center}
\caption{Metallicity of SDSS galaxies compared to the metallicity estimated through the FMR.
The solid line shows the one to one relation.}
\label{FMR}
\end{figure}

To compare how accurately the FMR can reproduce metallicity we follow \citet{Yates12}, since they use the same SDSS measurements of metallicity, SFR, and \Ms\ as in the current work. We used Eq.~(1) of \citet{Yates12} to estimate
$\mu_\alpha$=log(\Ms) $- \ \alpha$ log(SFR),  with $\alpha=0.19$. We then estimated metallicity
using Eq.~(2) of the same paper, given by 

{ \small \begin{equation}\label{FMR_EqY}
\rm 12+log(O/H)_{FMR} =43.447-12.193x+1.3728x^2-0.04985x^3,
\end{equation}} with $x=\mu_\alpha$. The comparison between the metallicity obtained through the FMR and the original value gives $\chi_{red}^2=1.06$. (Fig.~\ref{FMR}).


\subsection{FP and FMR as a function of signal to noise}

\begin{figure*}[t]
\begin{center}
\includegraphics[scale=0.7]{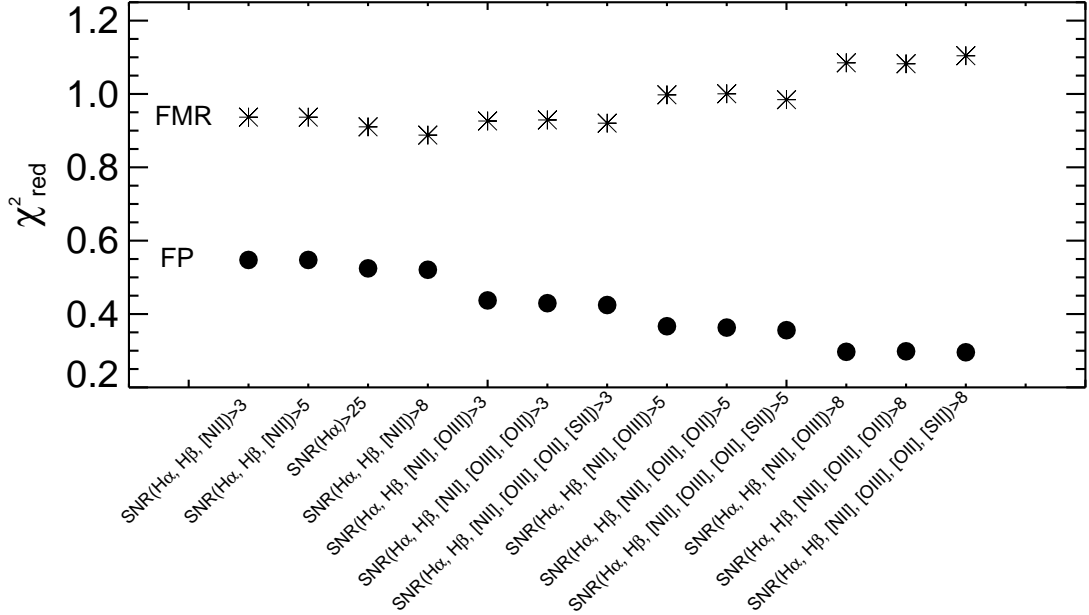}
\end{center}
\caption{ $\chi^2_{red}$ for the FP and FMR as a function of different values and combinations of SNR.}
\label{ChiSquareFPyFMR}
\end{figure*}

Here we explore the accuracy of the FP and FMR  with metallicity estimates based on emission lines selected at a variety of SNR thresholds. We estimate the $\chi^2_{red}$ using equation \ref{FPEq} to determine the \Ms\ and equation \ref{FMR_EqY} to determine metallicity for different values of SNR and for different combinations of emission lines (Fig. \ref{ChiSquareFPyFMR}).

Galaxies with a SNR(\Ha, \Hb, \NII)$>$3, SNR(\Ha, \Hb, \NII)$>$5, SNR(\Ha)$>$25, and  SNR(\Ha, \Hb, \NII)$>$8 show very similar $\chi^2_{red}$ for the FP and FMR.

For the FP, including galaxies with lines having low SNR increases the scatter thus giving higher $\chi^2_{red}$ values. Increasing the SNR in different lines decreases the dispersion giving lower $\chi^2_{red}$ values. The $\sigma$ of the residuals in the FP gradually decreases when the SNR increases going from $\sigma$=0.5 dex for a SNR(\Ha, \Hb, \NII)$>$3 to $\sigma$=0.29 dex for  a SNR(\Ha, \Hb, \NII, \OIII, \OII, \SII)$>$3

For the FMR, in contrast, restricting the sample to include only galaxies having lines of high SNR has the opposite effect, increasing the  $\chi^2_{red}$. When the SNR is increased, the $\sigma$ of the residuals also increases, going from $\sigma$=0.93 for a SNR(\Ha, \Hb, \NII)$>$3  to $\sigma$=1.06 for  a SNR(\Ha, \Hb, \NII, \OIII, \OII, \SII)$>$3.

The choice of threshold for the SNR of emission lines does not significantly change the metallicity ranges in our sample. It does, however, shift the locus of our sample $\sim$0.1 dex towards higher metallicity values when the extreme cases of Fig. \ref{ChiSquareFPyFMR} are compared.

\subsection{Results}

The summary in Table~\ref{TableChi} indicates that the fit that best minimizes $\chi_{red}^2$ for \Ms, $Z$, and SFR, is the plane corresponding to the regression on \Ms, the FP. Also, this plane is the only one that maintains linearity over the full range of values between the observed and the estimated
values in each projection.  At any SNR of different combinations of emission lines, the FP always gives the lowest $\chi_{red}^2$, indicating that this is the representation of the distribution that best minimises the variance.


Empirically, we can imagine that the \Ms$-Z$, \Ms$-$SFR, and $Z-$SFR relationships are the faces of this 3D distribution. The relationship that shows the highest dispersion (\S\,\ref{Z-SFR}) is the $Z-$SFR \citep{Lara10a, Lara10b}, which means that this relation is close to the face-on orientation of the 3D distribution (top panels of Fig.~\ref{Cubos3D}).

While \Ms\ correlates with both SFR and metallicity (the well known \Ms$-Z$ and \Ms$-$SFR relationships), the SFR does not  strongly correlate with metallicity. Therefore,  SFR and $Z$ are the least correlated variables, and a linear combination of these two parameters is enough to explain all three variables.

\section{The $Z-$SFR relation}\label{Z-SFR}

We turn now to a discussion of the $Z-$SFR relation, and the impact of taking the medians of the data in bins
defined in different order. We have just described the $Z-$SFR relation as being close to the face on view of the FP,
and the correlation between $Z$ and SFR is not intrinsically tight. In consequence,
any fit will suffer from a high degree of intrinsic scatter in the data around the fit.

This statement can be quantified using the Pearson correlation coefficient, which is a test of correlation between two variables, and is defined as $c= (\sigma^2_a$ $\times$ $\sigma^2_b$) / ($\sigma_a$ $\times$ $\sigma_b$). The quantity $c$ is a scalar in the interval $[-1.0, 1.0]$, where $-1.0$ and $1.0$ indicate a negative or positive perfect fit, respectively, while values close to $0.0$ would indicate a poor correlation. Applying this test to our relationships, we obtain $c=0.72$ for the \Ms$-Z$ relation, $c=0.76$ for the  \Ms$-$SFR relation, and $c=0.48$ for the $Z-$SFR relation. Therefore, of our three relationships, the $Z-$SFR  is the relation that shows the highest dispersion.

\begin{figure*}
 \centering
\includegraphics[scale=0.62]{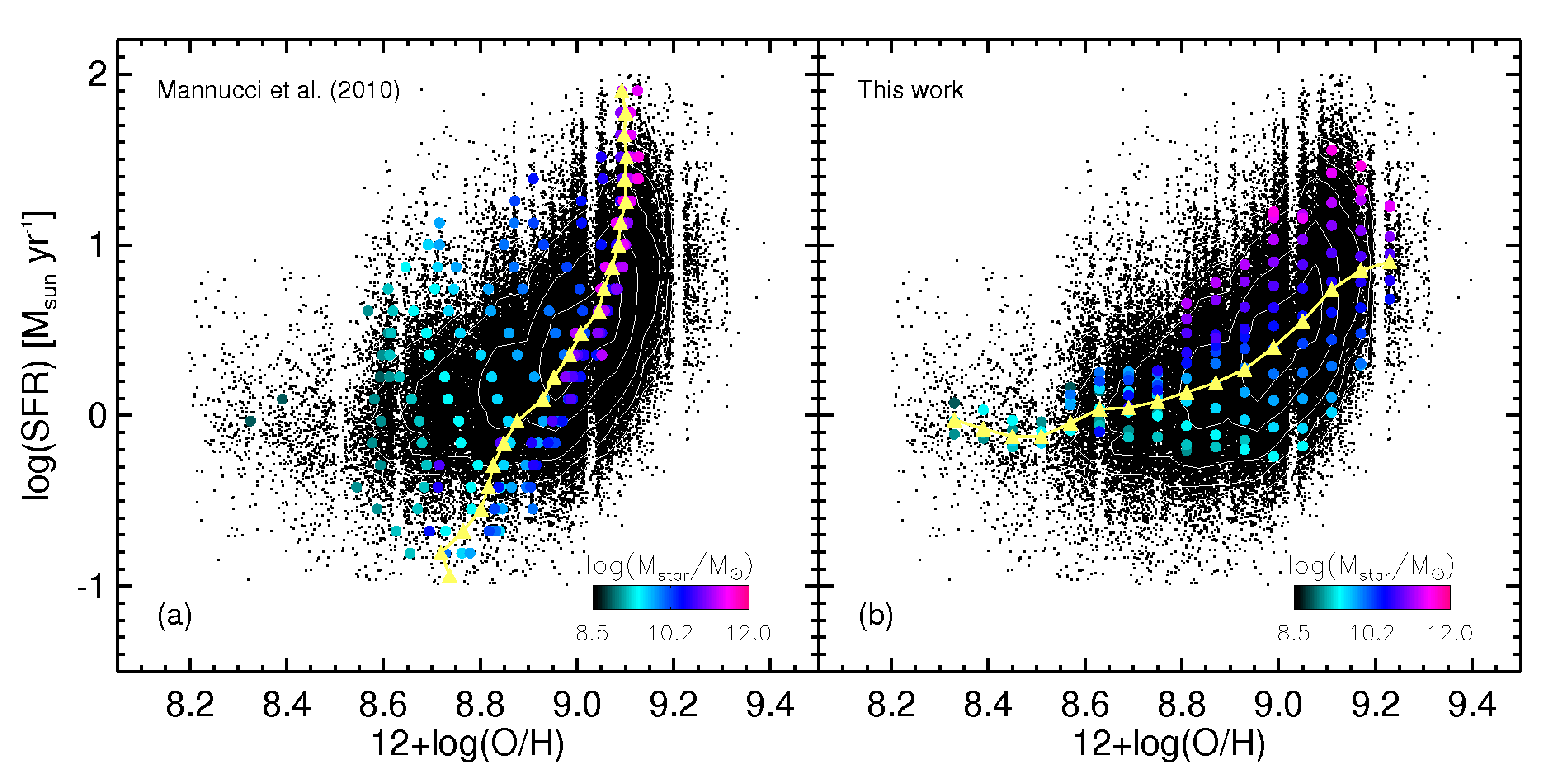}
\caption{(a) Filled coloured circles correspond to the $Z-$SFR relation binned following \citet{Mannucci10}, who took bins in $\log$(SFR) and estimated the median \abox\ per bin for different ranges of $\log (M_{\star} / M_{\sun})$, as shown in the color bar. (b) The same data, but now taking bins of \abox\ and estimating the median $\log$(SFR) using the same mass ranges as in (a). Black dots and white contours correspond to the SDSS sample. Yellow triangles show (a) the median metallicity in SFR bins, or (b) the median SFR in metallicity bins, considering the whole sample.}
\label{MetSFR}
\end{figure*}

Taking $Z$ as the key quantity, \citet{Mannucci10} studied the \Ms\ dependence of the $Z-$SFR relation by estimating the median $Z$ in bins of \Ms\ and SFR. This procedure can be thought as the projection of the FMR on the \Ms-SFR face of the 3D distribution. The result is shown in Fig.~\ref{MetSFR}a.  Although this relation has a high scatter, there is a tendency for the SFR generally to increase with $Z$. Fig.~\ref{MetSFR}a shows, however,
that binning as a function of mass reveals a more subtle effect, with the resulting median values showing
opposing trends depending on the mass selected. While the median metallicity increases with increasing
SFR for $\log($\Ms$) \gtrsim10.5$, it decreases with increasing SFR for $\log($\Ms$) \lesssim10.5$ (Fig.~\ref{MetSFR}a).

The mass dependence of the  $Z-$SFR relation can alternatively be explored by binning in a different order.
For every \Ms\ bin, we can estimate the median SFR in metallicity bins. Fig.~\ref{MetSFR}b clearly shows that this
binning order gives an apparently different result. For every \Ms, the median SFR is almost flat or slightly increasing with metallicity. There is a crucial distinction to be made here, which is one of correct interpretation. In the former case,
the median  metallicity has been estimated for a given \Ms\ and SFR, while in the latter, the median  SFR
has been estimated for a given \Ms\ and $Z$. It is true to say from the former approach that, at a given mass,
as SFR increases, the median  metallicity either increases (high-mass) or decreases (low-mass). It is also
true to say from the latter approach that, at a given mass, as metallicity increases, the median  SFR
either increases (high-mass) or stays relatively constant (low-mass). These statements are not inconsistent
with each other. The confusion arises when trying to impose an interpretation that is inconsistent with
the motivation for binning in a particular order. It would not be true to conclude from Fig.~\ref{MetSFR}a,
for example, that the median SFR decreases with increasing metallicity (low-mass) or increases
with increasing metallicity (high-mass), since the median SFR has not been calculated here.

We emphasize here that when fitting a relation to a distribution of data, the full dataset should be used.
Fitting only to a representation of the full dataset that is derived from the median values of one parameter
in bins of the others will clearly result in different surfaces being derived, and will depend on the choice of
the parameter for which the median is estimated.

A detailed study to tease apart the complex interplay of \Ms, $Z$, SFR, and SSFR that exploits this use of
different binning order for all the relationships will be presented in \citet[][]{Lara13}.

\section{Discussion and conclusions}\label{conc}

The use of a reliable metallicity estimator is crucial when analyzing the SFR dependence of the \Ms$-Z$ relation as shown in \citet{Yates12}. For this reason, we recommend that the estimator of N06 be used with caution, and limited to the range (12+log(O/H)$<$8.8) where the saturation of the N2 parameter is not a problem.

The emission-line galaxy spectra from SDSS are high quality, and measurements for many emission lines are available, making it possible to determine the gas-phase metallicity more robustly by applying  techniques which consider the ionization degree of the gas. Examples of these methods are  \citet{McGaugh91,KD02,KK04} and \citet{Tremonti04} (which are based on photoionization models) and \citet{P01a,P01b,PT05} and \citet{PVT10} (which rely on datasets for which $Z$ is known using the \Te\ method). 

We analyzed the 3D distribution of \Ms, Z, and SFR using three different approaches: $(i)$ fitting a plane using PCA, $(ii$) fitting a plane through regression \citep{Lara10a}, and $(iii)$ binning in SFR and \Ms\ to obtain the median metallicity of each bin \citep{Mannucci10}. For the five methods used, we estimated the $\chi_{red}^2$ as a measure of goodness of fit (Table \ref{TableChi}). We find that the best representation of the data is the plane defined by regression on \Ms.

We compare the \citet{Mannucci10}  surface (the FMR) and the \citet{Lara10a} Fundamental Plane (FP),  and demonstrate that the best representation of the data corresponds to a plane. While PCA  does not provide the best representation, it does demonstrate that the 3D distribution can be adequately represented
in two dimensions. The \Ms$-Z$, \Ms$-$SFR, and $Z-$SFR relationships are then projections of this plane.


We also highlight that the plane found by the regression fit on \Ms\
is not developed as a new technique for stellar mass estimation. Rather,
this approach is primarily aimed at identifying the most concise
representation of \Ms, $Z$ and SFR in order to facilitate more detailed
exploration of the interplay between these properties of galaxies.
Nevertheless, in cases when more robust techniques to estimate the stellar
mass \citep[e.g.,][]{Taylor11} are not available, the use of the FP could
be used to estimate the stellar mass, being careful to take into account
the metallicity and SFR uncertainties.

Our analysis of the $Z-$SFR relation with the two approaches toward binning the data highlights a crucial
need for caution in interpretation when exploring distributions represented as median values. 
The inappropriate interpretation of such results will lead to apparently contradictory conclusions,
depending on the binning order used. Furthermore, presenting medians in bins as a three-dimensional
distribution will lead to differing representations of the data, depending on the binning order chosen.

The SFR of a galaxy relates to the amount of gas currently being converted into stars, and correlates with the
current mass in stars, while metallicity is a measure of the number of times that the gas has
been reprocessed by stars, and also correlates with the current mass in stars in a galaxy.
The fact that we can represent  \Ms\ as a linear combination of SFR and metallicity suggests that the
stellar mass of a galaxy can be thought as the rate at which a galaxy is currently forming stars (SFR),
plus a measure of the star formation history, here represented by the metallicity ($Z$),
corresponding to the amount of reprocessing of the gas by past stellar generations.
The SF history and current SFR of a galaxy are closely linked to \Ms.

There is now an abundance of high quality spectroscopic measurements from large surveys such as the SDSS and the Galaxy And Mass Assembly (GAMA) survey \citep[][]{Driver11}. These resources provide the means to calculate robust metallicity estimators for significant numbers of galaxies. Managing the measurements from this growing data volume brings its own challenges, as well as the opportunity to explore scaling relations and broad population properties for statistically robust samples that can be divided into well-defined subsets in many different ways. Approaching these challenges and opportunities in the most robust way possible, by using the most accurate measurements available, will ensure that the most reliable scientific understanding of galaxy evolution can be produced.





\begin{acknowledgements}

We warmly thank the referee for extensive comments that have led to significant improvement in this paper.
We thank Mercedes Moll\'a, C\'esar Esteban, Alessandro Ederoclite, and  D\'amaso E. Chicharro-Mart\'{\i}nez for useful comments.
%
M. A. Lara-L\'opez thanks  the $``$Summer School in Statistics for Astronomers$"$, Center for Astrostatistics, PennState, especially to Arnab Chakraborty for invaluable tutorials on $``$R$"$ and PCA.

The work uses Sloan Digital Sky Survey (SDSS) data. Funding for the SDSS and SDSS-II was provided by the Alfred P. Sloan Foundation, the Participating Institutions, the National Science Foundation, the U.S. Department of Energy, the National Aeronautics and Space Administration, the Japanese Monbukagakusho, the Max Planck Society, and the Higher Education Funding Council for England. The SDSS was managed by the Astrophysical Research Consortium for the Participating Institutions. 
This research has made use of the SAO/NASA Astrophysics Data System Bibliographic Services (ADS).
\end{acknowledgements}


%
%
%
%
%
%
%
%
%


\begin{thebibliography}{14}


\bibitem[\protect\citeauthoryear {Abazajian et al.}{2009}]{Abaza09} 
Abazajian, K.~N., Adelman-McCarthy, J.~K., Ag{\"u}eros, M.~A., et al.\ 2009, ApJS, 182, 543 

 \bibitem[\protect\citeauthoryear {Adelman-McCarthy et al.}{2007}]{Adelman07} 
 Adelman--McCarthy, J. K., Ag{\"u}eros, M.~A., Allam, S.~S., et al. 2007, ApJs, 172, 634

\bibitem[\protect\citeauthoryear {Baldwin, Phillips \& Terlevich}{1981}]{Baldwin81} 
Baldwin J., Phillips M., Terlevich R., 1981, PASP, 93, 5 (BPT)

\bibitem[\protect\citeauthoryear{Bresolin et al.}{Bresolin et al.}{2009}]{Bresolin09}
Bresolin, F., Gieren, W., Kudritzki, R-P., Pietrzy\'nski, G., Urbaneja, M.A. \& Carraro, G. 2009, ApJ, 700, 309

\bibitem[\protect\citeauthoryear {Brinchmann et al.}{2004}]{Brinchmann04} 
Brinchmann, J., Charlot, S., White, S. D. M., et al. 2004, MNRAS, 351, 1151

\bibitem[\protect\citeauthoryear {Charlot \& Longhetti}{2001}]{Charlot01} 
Charlot, S., \& Longhetti, M.\ 2001, MNRAS, 323, 887 

\bibitem[\protect\citeauthoryear{Dav\'e \& Oppenheimer}{Dav\'e \& Oppenheimer}{2007}]{Dave07}
Dav\'e, R. \& Oppenheimer, B.D. 2007, MNRAS, 374, 427

\bibitem[\protect\citeauthoryear{De Lucia, Kauffmann \& White}{De Lucia et al.}{2004}]{DeLucia04}
De Lucia, G., Kauffmann, G. \& White, S.D.M. 2004, MNRAS, 374, 323

\bibitem[\protect\citeauthoryear{De Rossi, Tissera \& Scannapieco}{De Rossi et al.}{2006}]{DeRossi06}
De Rossi, M.E., Tissera, P.B., \& Scannapieco, C. 2006, MNRAS, 374, 323


\bibitem[\protect\citeauthoryear{Dors et al.}{2011}]{Dors11} Dors, O.~L., Jr., Krabbe, A., H{\"a}gele, G.~F., \& P{\'e}rez-Montero, E.\ 2011, MNRAS, 415, 3616 


\bibitem[\protect\citeauthoryear {Driver et al.}{2011}]{Driver11} Driver, S.~P., Hill, 
D.~T., Kelvin, L.~S., et al.\ 2011, MNRAS, 413, 971 



\bibitem[\protect\citeauthoryear {Ellison et al.}{2008}]{Ellison08} 
Ellison, S.~L., Patton, D.~R., Simard, L., et al. 2008, ApJL, 672, L107 


\bibitem[\protect\citeauthoryear{Esteban et al.}{Esteban et al.}{2009}]{Esteban09}
Esteban, C., Bresolin, F., Peimbert, M., Garc\'{\i}a-Rojas, J., Peimbert, A. \& Mesa-Delgado, A. 2009, ApJ, 700, 654


\bibitem[\protect\citeauthoryear{Foster et al.}{2012}]{Foster12} Foster, C., Hopkins, A.~M., Gunawardhana, M., et al.\ 2012, A\&A, 547, A79 


\bibitem[\protect\citeauthoryear{Garc\'{\i}a-Rojas \& Esteban}{Garc\'{\i}a-Rojas \& Esteban}{2007}]{GRE07}
Garc\'{\i}a-Rojas, J. \& Esteban, C., 2007, ApJ, 670, 457


\bibitem[\protect\citeauthoryear {Hopkins et al.}{2003}]{Hopkins03} Hopkins, A.~M., Miller, C.~J., Nichol, R.~C., et al.\ 2003, ApJ, 599, 971 

\bibitem[\protect\citeauthoryear {Kauffmann et al.}{2003}]{Kauf03} 
Kauffmann, G., Heckman, T. M., Tremonti, C., et al.\ 2003, MNRAS, 346, 1055 

\bibitem[\protect\citeauthoryear{Kewley \& Dopita}{Kewley \& Dopita}{2002}]{KD02}
Kewley, L.J. \& Dopita, M.A. 2002, ApJS, 142, 35
\bibitem[\protect\citeauthoryear{Kewley \& Ellison}{Kewley \& Ellison}{2008}]{KE08}
Kewley, L.J., \& Ellison, S.E. 2008, ApJ, 681, 1183
\bibitem[\protect\citeauthoryear {Kewley et al.}{2005}]{Kewley05} 
Kewley, L.~J., Jansen, R.~A., \& Geller, M.~J.\ 2005, PASP, 117, 227 

\bibitem[\protect\citeauthoryear{Kobulnicky \& Kewley}{Kobulnicky \& Kewley}{2004}]{KK04} 
Kobulnicky H.~A. \& Kewley L.~J. 2004, ApJ, 617, 240 

\bibitem[\protect\citeauthoryear{Lara-L{\'o}pez et al.}{2010a}]{Lara10a} 
Lara-L{\'o}pez, M.~A., Cepa, J., Bongiovanni, A., et al.\ 2010a, A\&A, 521, L53 

\bibitem[\protect\citeauthoryear {Lara-L{\'o}pez et al.}{2010b}]{Lara10b} 
Lara-L{\'o}pez, M.~A., Bongiovanni, A., Cepa, J., et al.\ 2010b, A\&A, 519, A31

\bibitem[\protect\citeauthoryear {Lara-L{\'o}pez et al.}{2013}]{Lara13} 
Lara-L{\'o}pez, M.~A., Hopkins, A. M., L\'opez-S\'anchez, A. R. et al.\ 2013, MNRAS, (submitted)


\bibitem[\protect\citeauthoryear{Lequeux et al.}{1979}]{lequeux79} 
Lequeux, J., Peimbert, M., Rayo, J. F., Serrano, A., \& Torres-Peimbert, S. 1979, A\&A, 80, 155


\bibitem[\protect\citeauthoryear{Liang et al.}{2006}]{Liang06} 
Liang, Y. C., Yin, S. Y., Hammer, F., et al. 2006, ApJ, 652, 257

\bibitem[\protect\citeauthoryear{L\'opez-S\'anchez \& Esteban}{L\'opez-S\'anchez \& Esteban}{2009}]{LSE09} 
L\'opez-S\'anchez, \'A.R. \& Esteban, C. 2009, A\&A, 508, 615

\bibitem[\protect\citeauthoryear{L\'opez-S\'anchez \& Esteban}{L\'opez-S\'anchez \& Esteban}{2010}]{LSE10b} 
L\'opez-S\'anchez, \'A.R. \& Esteban, C. 2010, A\&A, 517, 85

\bibitem[\protect\citeauthoryear{L\'opez-S\'anchez et al.}{L\'opez-S\'anchez et al.}{2011}]{LS+IC10+11} 
L\'opez-S\'anchez, \'A.R., Mesa-Delgado, A., L\'opez-Mart\`{\i}n, L \& Esteban, C. 2011, MNRAS, 411, 2076


\bibitem[\protect\citeauthoryear{L\'opez-S\'anchez et al.}{L\'opez-S\'anchez et al.}{2007}]{LSEGRPR07}
L\'opez-S\'anchez, \'A.R., Esteban, C., Garc\'{\i}a-Rojas, J., Peimbert, M. \& Rodr\'{\i}guez, M. 2007, ApJ, 656, 168



\bibitem[\protect\citeauthoryear{L\'opez-S\'anchez et al.}{L\'opez-S\'anchez et al.}{2012}]{LSDK+12}
L\'opez-S\'anchez, \'A.R., Dopita, M.A., Kewley, L.J.- Zahid, H.J., Nicholls, D.C. \& Scharw\"achter, J. 2012, MNRAS, in press


\bibitem[\protect\citeauthoryear{Mannucci et al.}{2010}]{Mannucci10} Mannucci, F., Cresci, G., Maiolino, R., Marconi, A., \& Gnerucci, A.\ 2010, MNRAS, 408, 2115 


\bibitem[\protect\citeauthoryear{Mart{\'{\i}}nez-Serrano et al.}{2008}]{Martinez10} Mart{\'{\i}}nez-Serrano, F.~J., Serna, A., Dom{\'{\i}}nguez-Tenreiro, R., \& Moll{\'a}, M.\ 2008, MNRAS, 388, 39 


\bibitem[\protect\citeauthoryear{McGaugh}{McGaugh}{1991}]{McGaugh91}
McGaugh, S.S. 1991, ApJ, 380, 140



\bibitem[\protect\citeauthoryear{Mesa-Delgado \& Esteban}{Mesa-Delgado \& Esteban}{2010}]{Mesa-Delgado10}
Mesa-Delgado, A., \& Esteban, C.\ 2010, MNRAS, 405, 2651



\bibitem[\protect\citeauthoryear{Moll{\'a} \& D{\'{\i}}az}{2005}]{mol05} 
Moll{\'a} M., D{\'{\i}}az A.~I., 2005, MNRAS, 358, 521 

\bibitem[\protect\citeauthoryear{Moustakas et al.}{Moustakas et al.}{2010}]{Moustakas+10}
Moustakas, J., Kennicutt, R.C., Jr., Tremonti, C. A., Dale, D. A., Smith, J.-D. T. \& Calzetti, D. 2010, ApJS, 190, 233

\bibitem[\protect\citeauthoryear{Nagao et al.}{Nagao, Maiolino \& Marconi}{2006}]{Nagao06}
Nagao, T., Maiolino, R. \& Marconi, A. 2006, A\&A, 459, 85

\bibitem[\protect\citeauthoryear{Peimbert et al.}{2007}]{Peimbert07}
Peimbert, M., Peimbert, A., Esteban, C., et al.\ 2007, RMxAC, 29, 72


\bibitem[\protect\citeauthoryear{Pettini \& Pagel}{Pettini \& Pagel}{2004}]{PP04}
Pettini, M. \& Pagel, B.E.J. 2004, MNRAS, 348, 59

\bibitem[\protect\citeauthoryear{Pilyugin}{Pilyugin}{2001a}]{P01a}
Pilyugin, L.S. 2001a, A\&A, 369, 594
\bibitem[\protect\citeauthoryear{Pilyugin}{Pilyugin}{2001b}]{P01b}
Pilyugin, L.S. 2001b, A\&A, 374, 412
\bibitem[\protect\citeauthoryear{Pilyugin \& Thuan}{Pilyugin \& Thuan}{2005}]{PT05}
Pilyugin, L.S. \& Thuan, T.X. 2005, ApJ, 631, 231
\bibitem[\protect\citeauthoryear{Pilyugin, V\'{\i}lchez \& Thuan}{Pilyugin et al.}{2010}]{PVT10}
Pilyugin, L.S., V\'{\i}chez, J.M. \& Thuan, T.X. 2010, ApJ, 720, 1738


\bibitem[\protect\citeauthoryear {Shlens}{Shlens}{2009}]{Shlens09}
Shlens, J., 2009, \emph{A Tutorial on Principal Component Analysis, Version 3.1}, 
New York University/Systems Neurobiology Laboratory, Salk 
Institute for Biological Studies

\bibitem[\protect\citeauthoryear {Strauss et al.}{2002}]{Strauss02} 
Strauss, M.~A., et al.\ 2002, AJ, 124, 1810 

\bibitem[\protect\citeauthoryear {Taylor et al.}{2011}]{Taylor11}
Taylor, E. N., Hopkins, 
A.~M., Baldry, I.~K., et al.\ 2011, MNRAS, 418, 1587


\bibitem[\protect\citeauthoryear{Tissera, De Rossi \& Scannapieco}{Tissera et al.}{2005}]{Tissera05}
Tissera, P.B., De Rossi, M.E., \& Scannapieco, C. 2005, MNRAS, 364, L38

\bibitem[\protect\citeauthoryear {Tremonti et al.}{2004}]{Tremonti04} Tremonti, C.~A., Heckman, T. M., Kauffmann, G., et al.\ 2004, ApJ, 613, 898 

\bibitem[\protect\citeauthoryear {Yates et al.}{2012}]{Yates12} Yates, R.~M., Kauffmann, G., \& Guo, Q.\ 2012, MNRAS, 422, 215


\bibitem[\protect\citeauthoryear{Yin et al}{Yin et al}{2007}]{Yin+07} 
Yin, S.Y., Liang, Y.C., Hammer, F., Brinchmann, J., Zhang, B., Deng, L.C. \& Flores, H., 2007, A\&A, 462, 535


\end{thebibliography}
\end{document}